\documentclass[aps,prl,twocolumn,superscriptaddress,nofootinbib]{revtex4-2}

\usepackage{amsmath,amssymb,bm,mathtools}
\usepackage{graphicx}
\usepackage{booktabs,array}
\usepackage[colorlinks=true,allcolors=blue]{hyperref}
\usepackage{microtype}
\usepackage{needspace} 

\newcommand{\dd}{\mathrm{d}}
\newcommand{\ii}{\mathrm{i}}
\newcommand{\Tr}{\operatorname{Tr}}

\begin{document}

\title{Atomistic theory of the $d$-wave phonon angular momentum splitter effect}

\author{Daniel A. Bustamante Lopez}
\email{daniel.bustamantelopez@unifr.ch}
\affiliation{Department of Physics, University of Fribourg, CH-1700 Fribourg, Switzerland}
\affiliation{Department of Applied Physics and Science Education, Eindhoven University of Technology, 5612 AP Eindhoven, Netherlands}

\author{Dominik M. Juraschek}
\email{dominik.juraschek@unifr.ch}
\affiliation{Department of Physics, University of Fribourg, CH-1700 Fribourg, Switzerland}
\affiliation{Department of Applied Physics and Science Education, Eindhoven University of Technology, 5612 AP Eindhoven, Netherlands}
\affiliation{Paul Scherrer Institut, CH-5232 Villigen PSI, Switzerland}

\date{\today}

\begin{abstract}
Phonons offer a bosonic route to angular-momentum transport in solids, but how their momentum-space textures manifest in transport remains unclear. Here we develop an atomistic theory of the phonon angular momentum splitter effect in an antiferromagnetic square lattice with checkerboard-modulated elastic couplings, analogous to the spin-splitter effect in $d$-wave altermagnets. We calculate the transport and accumulation of phonon angular momentum driven by a temperature gradient and find that the transverse response separates into Hall and $d_{xy}$-wave splitter contributions. The splitter contribution requires both the antiferromagnetic order and the checkerboard elastic modulation and exhibits a characteristic $d_{xy}$ angular dependence, whereas the Hall contribution is nearly isotropic. Their combination enables directional control of the edge phonon angular momentum accumulation. 
\end{abstract}

\maketitle

\section{Introduction}

Altermagnets are compensated magnets whose sublattices with opposite spins are connected by crystal-rotation symmetries, producing momentum-dependent nonrelativistic spin splitting despite vanishing net magnetization \cite{smejkal2022phase,smejkal2022landscape,jungwirth2026symmetry,jungwirth2026spintronics}. In $d$-wave altermagnets, the spin splitting forms a quadrupolar pattern in momentum space, enabling the spin-splitter effect: a longitudinal electric field drives spin-up and spin-down currents along different directions and generates a transverse spin current without relying on the conventional relativistic spin Hall mechanism \cite{naka2019spincurrent,gonzalezhernandez2021splitter,hallberg2025visualization,sigales2026splitter,sarkar2026extrinsic}. Spin-splitter torques and tilted spin currents have been reported in RuO$_2$-based heterostructures \cite{karube2022torque,bose2022tilted}.

Phonons provide an alternative bosonic route to angular-momentum transport, as circular or elliptical ionic motion carries phonon angular momentum (PAM) \cite{zhang2014pam,zhang2015chiral,zhu2018chiral,ueda2023quartz,ishito2023chiral,juraschek2025chiral}. Under a temperature gradient, PAM can be driven out of equilibrium, generating a PAM density through a phonon analogue of the Edelstein effect \cite{hamada2018gradient}, and, through the PAM Hall effect, a transverse PAM current with opposite accumulations at the sample edges, analogous to the spin Hall effect \cite{park2020pamhe,oh2025skew,bustamante2026atomistic}. PAM can further be induced by an electric field or a charge current \cite{hamada2020rotoelectric,yokoyama2025edelstein}. Experimental evidence includes the mechanical detection of thermally generated PAM in chiral tellurium, chirality-selective transport in quartz, and chiral-phonon-activated spin and orbital Seebeck effects \cite{zhang2025measurement,ohe2024quartz,kim2023spinseebeck,nabei2026orbitalseebeck}. 

In altermagnets, theoretical studies have predicted chiral and axial phonons, PAM Nernst responses, circular phonon dichroism, and momentum-dependent PAM textures \cite{rieger2026chiral,consiglio2026crsb,leon2026cuf2,wang2025alteraxial,li2026dichroism}, where the latter can also arise without magnetic order \cite{you2026structural}. Of particular relevance here, a phonon analogue of the spin-splitter effect has been proposed in $d$-wave altermagnets through magnon--phonon hybridization and orbital mechanisms \cite{bendin2026dwave,chatzichrysafis2026splitter}. These studies show that altermagnetic order can imprint its momentum-space symmetry on PAM, but how this texture manifests as a PAM current in real space and as a measurable edge accumulation in a finite sample remains unresolved.

\begin{figure}[!t]
    \centering
    \includegraphics[width=\linewidth]{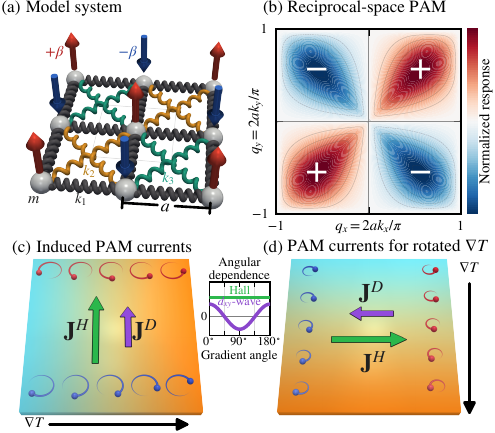}
\caption{PAM Hall and splitter responses in an antiferromagnetic square lattice with checkerboard elastic couplings. (a) The antiferromagnetic order couples to the lattice through alternating effective gyroscopic couplings $\pm\beta$, while axial ($k_1$) and checkerboard diagonal ($k_2,k_3$) springs provide the elastic interactions. (b) Checkerboard PAM texture in momentum space showing its $d_{xy}$ symmetry. (c,d) PAM currents from the Hall ($\mathbf{J}^{H}$) and $d_{xy}$-wave splitter ($\mathbf{J}^{D}$) effects for two orthogonal temperature gradients $\boldsymbol{\nabla}T$. Inset: angular dependence of the Hall (green) and $d_{xy}$-wave (purple) currents.}
\label{fig:pam_splitter_overview}
\end{figure}

Here, we develop an atomistic theory of the PAM splitter effect, illustrated in Fig.~\ref{fig:pam_splitter_overview}. We consider an antiferromagnetic square lattice in which the diagonal force constants alternate in a checkerboard pattern and the magnetic order couples directly to PAM. This minimal model supports a $d_{xy}$-wave PAM response alongside a PAM Hall contribution. Using the nonequilibrium steady-state covariance framework we developed for PAM transport in Ref.~\cite{bustamante2026atomistic}, we calculate the covariances of the atomic displacements and velocities, from which we obtain the local PAM, bond-resolved PAM current, and edge accumulation induced by a temperature gradient. We find that the transverse response separates into PAM Hall and $d$-wave components, with the latter changing sign upon rotating the temperature gradient by $90^\circ$. Their combination enables directional control of the transverse PAM current and edge accumulation, providing a signature of the $d$-wave PAM texture in real space.

\section{Model and transport formalism}
\label{sec:model}

We begin by describing a minimal model supporting a $d$-wave PAM texture. The model consists of a two-dimensional square lattice with axial nearest-neighbor springs of stiffness $k_1$ and diagonal springs alternating between $k_2$ and $k_3$ on neighboring plaquettes, as shown in Fig.~\ref{fig:pam_splitter_overview}(a). We write $k_2=\bar{k}(1-\delta)$ and $k_3=\bar{k}(1+\delta)$, so that $\delta$ measures the checkerboard contrast and $\delta\rightarrow-\delta$ interchanges the two elastic domains. The lattice carries antiferromagnetic order described by the staggered N\'eel parameter $N_z\eta_i$, with $\eta_i=\pm1$ and zero sum within the $2a\times2a$ unit cell. In the phonon dynamics, this spin--PAM coupling is represented by an effective sublattice-dependent gyroscopic coupling $\beta_i=g N_z\eta_i\equiv\beta\eta_i$, where $g$ is the effective spin--lattice coupling strength and $\beta$ has units of frequency \cite{fahrensohn2026phonomagnet,coh2023classification}. This effective coupling captures an interaction of the form $\mathbf{L}_i\cdot\mathbf{S}_i$ between phonon angular momentum and electron spin, which can arise from spin--orbit effects in magnetic materials \cite{garanin2015angular,ruckriegel2020angular,coh2023classification,ren2024adiabatic,juraschek2022giant,luo2023effective,geilhufe2022inertia,geilhufe2023electron,shabala2025axial,kahana2024magnonic}. Reversing the antiferromagnetic domain, $N_z\rightarrow-N_z$, therefore reverses $\beta$. This staggered velocity-dependent coupling can also be realized in actively driven gyroscopic lattices, which provide a classical mechanical analogue \cite{bendin2026dwave,nash2015gyroscopic}. 

The equation of motion for the in-plane displacement given by $\mathbf{u}_i=(u_{ix},u_{iy})^T$ is
\begin{equation}
m\ddot{\mathbf{u}}_i+2m\beta_i\mathbf{J}\dot{\mathbf{u}}_i+m\gamma_i\dot{\mathbf{u}}_i+\sum_{j\in\mathcal{N}_i}\boldsymbol{\Phi}_{ij}(\mathbf{u}_i-\mathbf{u}_j)=\boldsymbol{\xi}_i ,
\label{eq:langevin_main}
\end{equation}
where $m$ is the mass, $\mathbf{J}=\bigl(\begin{smallmatrix}0&-1\\1&0\end{smallmatrix}\bigr)$ rotates an in-plane vector by $90^\circ$, $\boldsymbol{\Phi}_{ij}$ is the force-constant matrix of bond $i-j$, $\gamma_i$ is the local damping rate, and $\boldsymbol{\xi}_i$ is the fluctuating force exerted by the thermal reservoir at site $i$. The Langevin reservoirs obey $\langle\xi_{i\mu}(t)\xi_{j\nu}(t')\rangle=2m\gamma_i k_{\mathrm B}T_i\,\delta_{ij}\delta_{\mu\nu}\delta(t-t')$, where $i,j$ label sites, $\mu,\nu$ label Cartesian components, and $T_i$ is the local reservoir temperature.

The mechanical PAM at site $i$ is given by $L_i^z=m(u_{ix}\dot{u}_{iy}-u_{iy}\dot{u}_{ix})$. To determine its momentum-space texture, we consider the periodic lattice and neglect damping and thermal noise in Eq.~\eqref{eq:langevin_main}. Applying Bloch's theorem to the $2a\times2a$ unit cell, we solve the resulting harmonic equations of motion to obtain the phonon frequencies and polarization vectors. The relative amplitudes and phases of the polarization components determine the ionic motion and hence the mechanical PAM of each phonon mode. We then isolate the checkerboard contribution by selecting the part of the mode PAM that changes sign under both $\beta\rightarrow-\beta$ and $\delta\rightarrow-\delta$. To leading order in these couplings, this contribution for phonon branch $n$ at momentum $\mathbf{k}$ is
\begin{equation}
L_{n\mathbf{k},\mathrm{cb}}^z=
\hbar\beta\delta\,F_n(\mathbf{k})
\sin(2k_xa)\sin(2k_ya),
\label{eq:dxy_texture_main}
\end{equation}
producing the $d_{xy}$ PAM texture shown in Fig.~\ref{fig:pam_splitter_overview}(b). $F_n(\mathbf{k})$ is even under $k_x\rightarrow-k_x$, $k_y\rightarrow-k_y$, and $k_x\leftrightarrow k_y$. Details of the derivation are given in Supplemental Material.

We determine the nonequilibrium steady state for a finite sample of $N_{\mathrm s}$ sites from the equal-time displacement and velocity covariances. To do so, we collect all displacements and velocities in the phase-space vector $\mathbf{z}=(\mathbf{u},\dot{\mathbf{u}})^T$. Eq.~\eqref{eq:langevin_main} for the full lattice can then be rewritten as
\[
\dot{\mathbf{z}}=\mathbf{A}\mathbf{z}+\boldsymbol{\zeta},
\qquad
\mathbf{A}=
\begin{pmatrix}
0&\mathbf{I}\\
-m^{-1}\mathbf{K}&-2\mathbf{B}-\boldsymbol{\Gamma}
\end{pmatrix},
\]
where $\boldsymbol{\zeta}=(0,m^{-1}\boldsymbol{\xi})^T$ represents the reservoir noise.  $\mathbf{K}$ is the full force-constant matrix assembled from the bond matrices $\boldsymbol{\Phi}_{ij}$ in Eq.~\eqref{eq:langevin_main}, with $\mathbf{K}_{ii}=\sum_{j\in\mathcal{N}_i}\boldsymbol{\Phi}_{ij}$ and $\mathbf{K}_{ij}=-\boldsymbol{\Phi}_{ij}$ for connected sites. $\mathbf{B}=\operatorname{diag}(\beta_1\mathbf{J},\ldots,\beta_{N_{\mathrm s}}\mathbf{J})$ contains the local gyroscopic couplings induced by the antiferromagnetic order, and $\boldsymbol{\Gamma}=\operatorname{diag}(\gamma_1\mathbf{I}_2,\ldots,\gamma_{N_{\mathrm s}}\mathbf{I}_2)$ contains the local damping rates. The equal-time covariance $\mathbf{C}=\langle\mathbf{z}\mathbf{z}^T\rangle$ is then found to satisfy $\dot{\mathbf{C}}=\mathbf{A}\mathbf{C}+\mathbf{C}\mathbf{A}^T+\mathbf{Q}$, where $\langle\boldsymbol{\zeta}(t)\boldsymbol{\zeta}^T(t')\rangle=\mathbf{Q}\delta(t-t')$.

To calculate the covariance induced by the temperature gradient in the nonequilibrium steady state, we set $\dot{\mathbf{C}}=0$ and write the reservoir temperatures as $T_i=T_0+\Delta T_i$. Expanding to linear order in $\Delta T_i$ about the uniform temperature $T_0$ yields the response $\Delta\mathbf{C}=\Delta\langle\mathbf{z}\mathbf{z}^T\rangle$,
\begin{equation}
\mathbf{A}\Delta\mathbf{C}+\Delta\mathbf{C}\mathbf{A}^T=-\Delta\mathbf{Q}.
\label{eq:lyapunov_main}
\end{equation}
For details, please see Supplemental Material. Since the reservoir noise $\boldsymbol{\zeta}$ acts only on the velocity components of the phase-space vector $\mathbf{z}$, the temperature-dependent noise response $\Delta\mathbf{Q}$ has only a velocity--velocity block,
\begin{equation}
(\Delta\mathbf{Q}_{vv})_{i\mu,j\nu}
=
\frac{2\gamma_i k_{\mathrm B}}{m}\,
\Delta T_i\,
\delta_{ij}\delta_{\mu\nu}.
\label{eq:noise_main}
\end{equation}
For a uniform temperature gradient, $\Delta T_i=(\mathbf{r}_i-\mathbf{r}_0)\cdot\nabla T$, Eqs.~\eqref{eq:lyapunov_main} and \eqref{eq:noise_main} give the exact classical linear response of the finite harmonic lattice. The mechanical PAM $L_i^z$ is bilinear in displacement and velocity, so its temperature-induced local accumulation follows directly from the displacement--velocity block of $\Delta\mathbf{C}$,
\begin{equation}
\Delta L_i^z=
m\left[
(\Delta\mathbf{C}_{uv})_{ix,iy}
-
(\Delta\mathbf{C}_{uv})_{iy,ix}
\right].
\label{eq:local_pam_main}
\end{equation}

We next turn to PAM transport through the lattice. Taking the cross product of $\mathbf{u}_i$ with Eq.~\eqref{eq:langevin_main} gives the local angular-momentum balance (see Supplemental Material),
\begin{equation}
\frac{d\lambda_i^z}{dt}
+\sum_{j\in\mathcal{N}_i}j^L_{i\rightarrow j}
=
\sum_{j\in\mathcal{N}_i}q^L_{ij}
+\tau_i^{\mathrm{bath}},
\label{eq:continuity_main}
\end{equation}
where $\lambda_i^z=L_i^z+m\beta_i|\mathbf{u}_i|^2$ is the canonical angular-momentum density. The quantity $2q^L_{ij}$ is the net torque exerted by bond $i-j$ on its two ionic endpoints; the elastic frame that fixes the bond direction experiences the opposite reaction torque.  $\tau_i^{\mathrm{bath}}=\mathbf{u}_i\times(\boldsymbol{\xi}_i-m\gamma_i\dot{\mathbf{u}}_i)$ is the torque exerted on site $i$ by the stochastic reservoir force and damping. $j^L_{i\rightarrow j}$ describes the angular momentum transferred from site $i$ to site $j$ and therefore defines the bond-resolved PAM current,
\begin{equation}
j^L_{i\rightarrow j}
=
\frac{1}{2}
(\mathbf{u}_i+\mathbf{u}_j)
\times
\boldsymbol{\Phi}_{ij}(\mathbf{u}_i-\mathbf{u}_j).
\label{eq:bond_current_main}
\end{equation}
Thus PAM can be transported between sites, exchanged with the elastic frame, or exchanged with the reservoirs. Explicit expressions for the bond torque and the global balance equation are given in Supplemental Material.

We now obtain the PAM current density by weighting each bond current by the bond displacement, 
\begin{equation}
J_a^L=
\frac{1}{\mathcal{A}}
\sum_{\langle ij\rangle}
(r_{j,a}-r_{i,a})
\langle j^L_{i\rightarrow j}\rangle
=
-\sum_b\sigma_{ab}^{L}\partial_bT ,
\label{eq:current_density_main}
\end{equation}
where $\mathcal{A}$ is the sample area, $a,b\in\{x,y\}$ denote the directions of the PAM current and temperature gradient, respectively, and $\sigma_{ab}^{L}$ is the PAM conductivity. The bond displacement $\mathbf{r}_j-\mathbf{r}_i$ resolves each microscopic transfer along the transport direction. Let $\mathbf{C}_{ij}^{(b)}=\partial\langle\mathbf{u}_i\mathbf{u}_j^T\rangle/\partial(\partial_bT)$ denote the $2\times2$ displacement-covariance block obtained by solving Eqs.~\eqref{eq:lyapunov_main} and \eqref{eq:noise_main} for a unit gradient along direction $b$. Substitution into Eq.~\eqref{eq:bond_current_main} then yields an explicit expression for the PAM conductivity,
\begin{equation}
\begin{aligned}
\sigma_{ab}^{L}
&= \frac{1}{2\mathcal{A}}
\sum_{\langle ij\rangle}
(r_{j,a}-r_{i,a}) \\
&\quad\times\mathrm{Tr}\!\left[
\mathbf{J}\boldsymbol{\Phi}_{ij}
\left(
\mathbf{C}_{ii}^{(b)}+\mathbf{C}_{ij}^{(b)}
-\mathbf{C}_{ji}^{(b)}-\mathbf{C}_{jj}^{(b)}
\right)\right].
\end{aligned}
\label{eq:conductivity_main}
\end{equation}
Eq.~\eqref{eq:conductivity_main} is the classical linear-response coefficient obtained from the bond-resolved PAM currents. For its equivalent quantum Green-function form, see Supplemental Material.

To characterize the PAM response of the finite sample, we define the torque absorbed at an edge $\alpha$ as $\tau_{\alpha}^{\mathrm{abs}}=\sum_i w_{\alpha i}\gamma_i\,\Delta L_i^z$, where $w_{\alpha i}$ selects the sites in the corresponding edge region. We then compare the torque absorbed at opposite edges and define its response to a temperature gradient,
\begin{equation}
\mathcal{T}_x=
\frac{\tau_{\mathrm{right}}^{\mathrm{abs}}-\tau_{\mathrm{left}}^{\mathrm{abs}}}{2\ell_y},
\mathcal{T}_y=
\frac{\tau_{\mathrm{top}}^{\mathrm{abs}}-\tau_{\mathrm{bottom}}^{\mathrm{abs}}}{2\ell_x},
G_{ab}^{L}=
-\frac{\partial\mathcal{T}_a}{\partial(\partial_bT)} ,
\label{eq:edge_response_main}
\end{equation}
where $\ell_x$ and $\ell_y$ are the sample dimensions. The resulting tensor $\mathbf{G}^{L}$ characterizes the PAM torque absorbed at opposite edges to the temperature gradient and provides the finite-sample counterpart of the bulk bond conductivity $\boldsymbol{\sigma}^{L}$. The two tensors have the same symmetry decomposition, but need not be numerically identical, because the angular-momentum balance in Eq.~\eqref{eq:continuity_main} also includes torques exchanged with the elastic frame and thermal reservoirs, as well as boundary contributions.

\begin{figure}[!t]
    \centering
    \includegraphics[width=0.98\columnwidth]{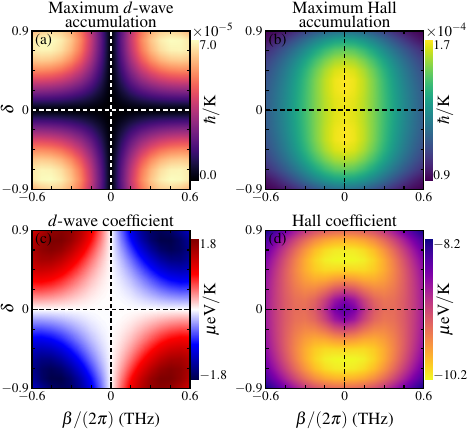}
\caption{Local PAM accumulation and edge-response coefficients as functions of the antiferromagnetic-order-induced coupling $\beta$ and checkerboard contrast $\delta$. (a,b) Maximum local $d$-wave and Hall PAM accumulations normalized by the applied temperature difference. (c,d) Corresponding $d$-wave and Hall edge-response coefficients.}
\label{fig:pam_beta_delta_map}
\end{figure}

To distinguish the $d_{xy}$-wave splitter response from the Hall one, we isolate the component that changes sign when either the antiferromagnetic domain ($\beta$) or the elastic checkerboard domain ($\delta$) is reversed. For either the bulk conductivity $\boldsymbol{\sigma}^{L}$ or the finite-sample edge-response tensor $\mathbf{G}^{L}$, denoted generically by $\mathbf{X}$, we obtain
{\small
\begin{equation}
\mathbf{X}_{\mathrm{cb}}(\beta,\delta)=
\tfrac{1}{4}
\bigl[
\mathbf{X}(\beta,\delta)-\mathbf{X}(-\beta,\delta)
-\mathbf{X}(\beta,-\delta)+\mathbf{X}(-\beta,-\delta)
\bigr].
\label{eq:domain_projection_main}
\end{equation}
}
The mirror symmetries of the square lattice eliminate the diagonal components of $\mathbf{X}_{\mathrm{cb}}$, while interchanging the $x$ and $y$ axes makes its off-diagonal components equal. We therefore identify the symmetric off-diagonal part with the $d_{xy}$-wave splitter coefficient $D$,
\begin{equation}
\mathbf{X}_{\mathrm{cb}}=
D
\begin{pmatrix}
0&1\\
1&0
\end{pmatrix},
\quad
D=\frac{X_{\mathrm{cb},xy}+X_{\mathrm{cb},yx}}{2},
\label{eq:D_definition_main}
\end{equation}
The PAM Hall coefficient $H$ follows from the antisymmetric part of the response, giving
\begin{equation}
\mathbf{X}=
\begin{pmatrix}
0&D-H\\
D+H&0
\end{pmatrix},
\quad
H=\frac{X_{yx}-X_{xy}}{2}.
\label{eq:combined_tensor_main}
\end{equation}
For an in-plane temperature gradient specified by $-\boldsymbol{\nabla}T=|\nabla T|(\cos\phi,\sin\phi)$, we further find that the longitudinal and transverse bulk currents satisfy
\begin{equation}
J_{\parallel}^{L}=D|\nabla T|\sin(2\phi),
\qquad
J_{\perp}^{L}=|\nabla T|\left[H+D\cos(2\phi)\right].
\label{eq:angular_response_main}
\end{equation}
The transverse $d$-wave current therefore reverses under a $90^\circ$ rotation of the temperature gradient, whereas the Hall contribution retains its sign.

We next examine the $d$-wave and Hall contributions to the PAM accumulation in real space. Let $\Delta L_x^z(x,y)$ and $\Delta L_y^z(x,y)$ denote the spatial distributions of induced local PAM generated by temperature profiles oriented along $x$ and $y$. We separate their $d$-wave and Hall parts as
\begin{equation}
\Delta L_D^z=
\frac{\Delta L_x^z+\mathcal{R}_{xy}\Delta L_y^z}{2},
\quad
\Delta L_H^z=
\frac{\Delta L_x^z-\mathcal{R}_{xy}\Delta L_y^z}{2},
\label{eq:map_decomposition_main}
\end{equation}
%
where $\mathcal{R}_{xy}$ exchanges spatial axes, $\mathcal{R}_{xy}f(x,y)=f(y,x)$, so that the distributions induced by orthogonal temperature profiles are compared at corresponding positions.

\section{Numerical results}
\label{sec:results}

In Fig.~\ref{fig:pam_beta_delta_map}, we show the $d$-wave and Hall contributions to the local PAM accumulation and edge-response coefficients as functions of the gyroscopic coupling $\beta$ and elastic checkerboard contrast $\delta$. For Fig.~\ref{fig:pam_beta_delta_map}(a,b), we apply a smooth temperature profile between hot and cold regions by setting the site-resolved reservoir temperatures to $T_i=T_0+\Delta T_i$, with the prescribed temperature changes $\Delta T_i$ entering the noise covariance in Eq.~\eqref{eq:noise_main}. Solving Eq.~\eqref{eq:lyapunov_main} gives the response $\Delta\mathbf{C}$, from which we calculate the local PAM accumulation using Eq.~\eqref{eq:local_pam_main}. We separate its $d$-wave and Hall parts according to Eq.~\eqref{eq:map_decomposition_main} and apply the domain projection in Eq.~\eqref{eq:domain_projection_main} to the $d$-wave component, while retaining the part even under both $\beta\rightarrow-\beta$ and $\delta\rightarrow-\delta$ for the Hall contribution. The maximum local accumulations are normalized by the applied temperature difference and expressed in units of $\hbar/\mathrm{K}$. 

For Fig.~\ref{fig:pam_beta_delta_map}(c,d), we use a unit linear temperature gradient and extract $D$ and $H$ from the edge-response tensor using Eqs.~\eqref{eq:edge_response_main}--\eqref{eq:combined_tensor_main}. Since these coefficients measure the edge torque per unit length in response to a temperature gradient, their units are energy per temperature, plotted in $\mu\mathrm{eV}/\mathrm{K}$. The splitter response vanishes on either axis, where one of the two ingredients is absent ($\beta=0$ or $\delta=0$). Away from these axes, the $d$-wave accumulation grows and $D$ reverses sign when either domain is reversed. Thus, the largest splitter signals occur when both $|\beta|$ and $|\delta|$ are large. By contrast, the Hall accumulation and coefficient remain finite at $\beta=\delta=0$ and vary only weakly across the parameter range, so the two contributions can be adjusted relative to one another.

\begin{figure}[!t]
    \centering
    \includegraphics[width=0.94\columnwidth]{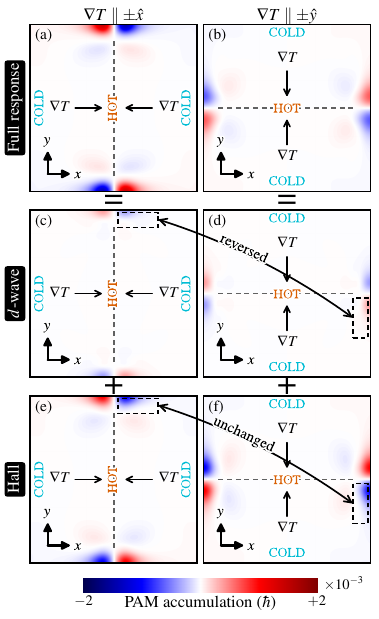}
    \caption{Real-space PAM accumulation generated by a central hot region for temperature profiles varying along $x$ (left column) and $y$ (right column). Red and blue indicate PAM accumulation along the positive and negative $z$ directions, respectively, corresponding to opposite senses of circular ionic motion. Top to bottom: full, $d$-wave, and Hall responses. At equivalent transverse edges, the $d$-wave contribution reverses sign after a $90^\circ$ rotation of the temperature-profile direction, whereas the Hall contribution retains its sign.}
    \label{fig:center_hot_local_pam}
\end{figure}

To illustrate the real-space PAM accumulation, we apply a temperature profile with a central hot region and colder opposite boundaries. We calculate the local PAM accumulation and decompose it using Eq.~\eqref{eq:map_decomposition_main}. In Fig.~\ref{fig:center_hot_local_pam}, we compare the PAM accumulation for temperature profiles oriented along $x$ (left column) and $y$ (right column). The top row shows the total PAM accumulation from the $d$-wave splitter and Hall contributions, which differs at the transverse edges for the two orientations of the temperature profile. The middle row isolates the $d$-wave contribution, which reverses sign at corresponding transverse edges after a $90^\circ$ rotation of the temperature profile. In contrast, the Hall contribution in the bottom row retains its sign. All maps use the same color scale for the PAM accumulation, allowing their magnitudes to be compared directly. Although the Hall response dominates for the parameters shown, the splitter contribution reinforces it for one orientation and partially cancels it for the other. This directional change in the total edge accumulation hence provides a clear signature of the PAM splitter effect.

\section{Discussion}
\label{sec:conclusion}

Measuring the edge PAM accumulation as the temperature gradient is rotated would provide a direct experimental test to distinguish the $d$-wave splitter response from the PAM Hall contribution. Reversing the antiferromagnetic order, for example through field cooling or current-induced switching \cite{liu2026mnte,bodnar2018mn2au}, provides a complementary test: the $d$-wave splitter response changes sign, whereas the Hall contribution remains unchanged. Strain offers an additional way to tune magnetic properties \cite{khodas2026strain}. Such directional control of PAM parallels the functionality of electronic spin splitters and suggests a route toward angular-momentum signal processing in chiral phononics.

Mechanical torque measurements could probe the predicted edge accumulation, building on the detection of thermally generated PAM in tellurium \cite{zhang2025measurement}. Related experiments have demonstrated transport of chiral phonons in quartz and their contributions to spin and orbital Seebeck effects \cite{ohe2024quartz,kim2023spinseebeck,nabei2026orbitalseebeck}. Because circular ionic motion can carry a phonon magnetic moment \cite{juraschek2019orbital,ren2021magneticmoment,juraschek2017dynamical,zhang2023gate,geilhufe2021induced,geilhufe2023electron}, spatially resolved magneto-optic Kerr or Faraday measurements could provide an indirect probe \cite{basini2024multiferroicity,luo2023effective,davies2024barnett}. X-ray scattering and ultrafast electron diffraction have resolved chiral or polarized phonon motion \cite{ueda2023quartz,tauchert2022polarized}, while circular phonon dichroism could probe chirality and altermagnetic order \cite{li2026dichroism}. 
%
%
Identifying materials that could host the effect is therefore an important next step and could be based on symmetry arguments \cite{wang2025alteraxial,you2026structural} or electronic $d$-wave candidate materials \cite{ohlendorf2026mnf2}. 


As a macroscopic alternative, a square array of motor-driven gyroscopes could provide a realization of the effect, building on gyroscopic lattices and topological mechanical metamaterials \cite{nash2015gyroscopic,mitchell2018transition,mitchell2018amorphous,susstrunk2015helical,susstrunk2016classification}. Alternating rotor circulation would realize the staggered couplings $\pm\beta$, while alternating diagonal springs would reproduce the checkerboard contrast. Spatially patterned stochastic driving could emulate the Langevin temperature profiles \cite{serragarcia2016stochastic}. The effect could be tested by measuring how the gyroscopes move while changing the driving direction, reversing their rotation, or swapping the diagonal springs.

\section*{Acknowledgments}

We thank Robin Neumann, Alexander Mook, Hannah Bendin, Quintin Meier, Andrea Urru, Xanthe Verbeek, and Shang Ren for useful discussions. This research was supported by the ERC Starting Grant CHIRALPHONONICS, No. 101166037.

\onecolumngrid
\clearpage
\setcounter{section}{0}
\setcounter{subsection}{0}
\setcounter{equation}{0}
\setcounter{figure}{0}
\setcounter{table}{0}
\renewcommand{\theequation}{S\arabic{equation}}
\renewcommand{\thefigure}{S\arabic{figure}}
\renewcommand{\thetable}{S\Roman{table}}
\renewcommand{\theHequation}{supplement.equation.\arabic{equation}}
\renewcommand{\theHfigure}{supplement.figure.\arabic{figure}}
\renewcommand{\theHtable}{supplement.table.\arabic{table}}
\renewcommand{\theHsection}{supplement.section.\arabic{section}}
\renewcommand{\theHsubsection}{supplement.section.\arabic{section}.subsection.\arabic{subsection}}
\begin{center}
{\large\bfseries Supplemental Material for \\ ``Atomistic theory of the $d$-wave phonon angular momentum splitter effect''}
\end{center}
\vspace{1em}

\section{Checkerboard lattice and equations of motion}

We consider a minimal two-dimensional square lattice in which each site carries two in-plane displacement coordinates. Nearest neighbours along the $x$ and $y$ directions are connected by central springs of stiffness $k_1$. The lattice also contains diagonal springs whose stiffness alternates between $k_2$ and $k_3$ from one plaquette to the next, producing an elastic checkerboard pattern. A collinear antiferromagnetic order is represented by a local gyroscopic coupling whose sign alternates between the two magnetic sublattices. The gyroscopic couplings sum to zero within the magnetic unit cell, while the staggered N\'eel order remains finite. The gyroscopic pattern is therefore compensated: its uniform component vanishes within one magnetic unit cell, but its staggered component remains nonzero.

We describe the dynamics of the lattice with the Langevin equation
\begin{equation}
 m\ddot{\mathbf u}_i
 +2m\beta_i\mathbf{J}\dot{\mathbf u}_i
 +m\gamma_i\dot{\mathbf u}_i
 +\sum_{j\in\mathcal N_i}\boldsymbol{\Phi}_{ij}(\mathbf u_i-\mathbf u_j)
 =\boldsymbol\xi_i,
 \label{eq:eom_site}
\end{equation}
where $\mathbf u_i=(u_{ix},u_{iy})^T$ is the displacement of site $i$, $m$ is the mass, $\mathcal N_i$ contains the sites connected to $i$, and $\boldsymbol{\Phi}_{ij}$ is the force-constant matrix of the bond $i-j$. The antisymmetric matrix
\begin{equation}
 \mathbf{J}=\begin{pmatrix}0&-1\\1&0\end{pmatrix},
 \qquad
 \mathbf{J}^T=-\mathbf{J},
 \label{eq:J}
\end{equation}
rotates an in-plane vector by $90^\circ$. The term $2m\beta_i\mathbf{J}\dot{\mathbf u}_i$ is therefore the local gyroscopic term associated with the staggered magnetic order. The coefficient $\gamma_i$ is the damping rate, and $\boldsymbol\xi_i$ is the fluctuating force exerted by a reservoir at temperature $T_i$. In the classical Langevin limit, $\boldsymbol\xi_i$ is a zero-mean Gaussian white-noise force satisfying
\begin{equation}
 \left\langle
 \xi_{i\mu}(t)\xi_{j\nu}(t')
 \right\rangle
 =
 2m\gamma_i k_{\mathrm B}T_i\,
 \delta_{ij}\delta_{\mu\nu}\delta(t-t'),
 \label{eq:classical_noise}
\end{equation}
where $i$ and $j$ label lattice sites, while $\mu$ and $\nu$ label Cartesian components.

We parametrize the coupling to the antiferromagnetic order as
$\beta_i=g N_z\eta_i\equiv\beta\eta_i$. Here $N_z$ is the signed amplitude of the collinear N\'eel order, $\eta_i=\pm1$ identifies the magnetic sublattice, and $g$ is an effective spin--lattice coupling that converts the order parameter into a gyroscopic rate. The normalization of $N_z$ can be absorbed into $g$, and $\beta=g N_z$ has dimensions of frequency. Reversing the antiferromagnetic domain changes $N_z\to-N_z$ and therefore $\beta\to-\beta$.

The conservative part of Eq.~\eqref{eq:eom_site} follows from the Lagrangian
\begin{equation}
 \mathcal L
 =
 \sum_i
 \left[
 \frac{m}{2}|\dot{\mathbf u}_i|^2
 +\beta_iL_i^z
 \right]
 -\frac12\sum_{\langle ij\rangle}
 (\mathbf u_i-\mathbf u_j)^T
 \boldsymbol{\Phi}_{ij}
 (\mathbf u_i-\mathbf u_j),
 \label{eq:lagrangian}
\end{equation}
where the first term is the kinetic energy, the second term couples the local mechanical angular momentum to the staggered antiferromagnetic order, and the final term is the elastic energy stored in the springs. The local mechanical angular momentum is
\begin{equation}
 L_i^z
 =
 m(u_{ix}\dot u_{iy}-u_{iy}\dot u_{ix})
 =
 -m\mathbf u_i^T\mathbf{J}\dot{\mathbf u}_i.
 \label{eq:local_L}
\end{equation}
The derivatives entering the Euler--Lagrange equation are
\begin{equation}
 \frac{\partial\mathcal L}{\partial\dot{\mathbf u}_i}
 =
 m\dot{\mathbf u}_i
 +m\beta_i\mathbf{J}\mathbf u_i
 \equiv\mathbf p_i,
 \qquad
 \frac{\partial\mathcal L}{\partial\mathbf u_i}
 =
 -m\beta_i\mathbf{J}\dot{\mathbf u}_i
 -\sum_j\boldsymbol{\Phi}_{ij}(\mathbf u_i-\mathbf u_j),
 \label{eq:EL_derivatives}
\end{equation}
where $\mathbf p_i$ is the canonical momentum. Substituting these derivatives into
\begin{equation}
 \frac{\dd}{\dd t}
 \frac{\partial\mathcal L}{\partial\dot{\mathbf u}_i}
 -
 \frac{\partial\mathcal L}{\partial\mathbf u_i}
 =
 \boldsymbol\xi_i-m\gamma_i\dot{\mathbf u}_i
 \label{eq:EL_open}
\end{equation}
reproduces Eq.~\eqref{eq:eom_site}, including the nonconservative damping and reservoir forces.

\subsection{Four-site checkerboard cell and spring forces}

The smallest cell containing both the elastic checkerboard and the staggered gyroscopic pattern has dimensions $2a\times2a$. We label its four sites by
\begin{equation}
 \mathbf r_A=(0,0),\qquad
 \mathbf r_B=(a,0),\qquad
 \mathbf r_C=(0,a),\qquad
 \mathbf r_D=(a,a),
 \label{eq:basis}
\end{equation}
with Bravais vectors $(2a,0)$ and $(0,2a)$. The staggered sublattice signs are
\begin{equation}
 (\eta_A,\eta_B,\eta_C,\eta_D)
 =
 (+1,-1,-1,+1).
 \label{eq:eta_cell}
\end{equation}
They satisfy
$\eta_A+\eta_B+\eta_C+\eta_D=0$ and therefore
$\beta_A+\beta_B+\beta_C+\beta_D=0$. This is the compensation condition for the gyroscopic order: the cell-averaged gyroscopic coupling vanishes even though the staggered coupling remains finite.

A central spring of stiffness $k$ directed along the unit vector $\hat{\mathbf n}$ exerts on site $i$ the force
$-k\hat{\mathbf n}\hat{\mathbf n}^T(\mathbf u_i-\mathbf u_j)$
due to its relative displacement from the neighboring site $j$. Its force-constant matrix is therefore
$k\hat{\mathbf n}\hat{\mathbf n}^T$. The axial and diagonal bonds are described by
\begin{align}
 \boldsymbol{\Phi}_x
 &=
 k_1
 \begin{pmatrix}
 1&0\\
 0&0
 \end{pmatrix},
 &
 \boldsymbol{\Phi}_y
 &=
 k_1
 \begin{pmatrix}
 0&0\\
 0&1
 \end{pmatrix},
 \label{eq:Phi_xy}\\
 \boldsymbol{\Phi}_+(k)
 &=
 \frac{k}{2}
 \begin{pmatrix}
 1&1\\
 1&1
 \end{pmatrix},
 &
 \boldsymbol{\Phi}_-(k)
 &=
 \frac{k}{2}
 \begin{pmatrix}
 1&-1\\
 -1&1
 \end{pmatrix}.
 \label{eq:Phi_diag}
\end{align}
The subscripts $+$ and $-$ refer to diagonal bonds parallel to $(1,1)$ and $(1,-1)$, respectively. Diagonal bonds belonging to alternating plaquettes carry stiffness $k_2$ or $k_3$ according to the checkerboard pattern. The complete periodic bond list for the four-site cell is given in Table~\ref{tab:bonds}.

We write the two diagonal stiffnesses as
\begin{equation}
 k_2=\bar k(1-\delta),\qquad
 k_3=\bar k(1+\delta),\qquad
 \bar k=\frac{k_2+k_3}{2},\qquad
 \delta=\frac{k_3-k_2}{k_3+k_2}.
 \label{eq:kbar_delta}
\end{equation}
The parameter $\bar k$ is the mean diagonal stiffness, while the dimensionless contrast $\delta$ measures the strength and domain of the elastic checkerboard modulation. Interchanging $k_2$ and $k_3$ changes $\delta\to-\delta$ and reverses the checkerboard domain. When $\delta=0$, the diagonal springs remain present with uniform stiffness $\bar k$, but the elastic checkerboard modulation is absent.

\section{Phonon spectrum and phonon angular momentum}

\subsection{Bloch equation for the checkerboard}

To obtain the phonon spectrum, we consider the isolated periodic lattice without damping or thermal reservoirs, so that $\gamma_i=0$ and $\boldsymbol\xi_i=0$. We use the Bloch displacement
\begin{equation}
 \mathbf u_{\ell\alpha}(t)=\boldsymbol\epsilon_\alpha(\mathbf k)
 \exp\!\left[\ii\mathbf k\cdot(\mathbf R_\ell+\mathbf r_\alpha)-\ii\omega t\right],
 \label{eq:bloch}
\end{equation}
where $\ell$ labels the $2a\times2a$ unit cell, $\alpha=A,B,C,D$ labels a site within the cell, $\mathbf R_\ell$ is a Bravais-lattice vector, $\mathbf r_\alpha$ is the basis position defined in Eq.~\eqref{eq:basis}, $\mathbf k$ is the crystal momentum, $\omega$ is the phonon frequency, and $\boldsymbol\epsilon_\alpha(\mathbf k)$ is the two-component polarization amplitude on site $\alpha$.

Substitution into Eq.~\eqref{eq:eom_site} gives
\begin{equation}
 \left[\mathbf{D}(\mathbf k)-\omega^2\mathbf{I}_8
 -2\ii\omega\beta\boldsymbol{\mathcal J}_{\mathrm{AF}}\right]
 \boldsymbol\epsilon_{n\mathbf k}=0,
 \qquad
 \boldsymbol{\mathcal J}_{\mathrm{AF}}
 =\operatorname{diag}(\eta_A\mathbf{J},\eta_B\mathbf{J},\eta_C\mathbf{J},\eta_D\mathbf{J})
 =\operatorname{diag}(\mathbf{J},-\mathbf{J},-\mathbf{J},\mathbf{J}).
 \label{eq:qep}
\end{equation}
Here $n$ labels the phonon branch and
$\boldsymbol\epsilon_{n\mathbf k}
=(\boldsymbol\epsilon_A,\boldsymbol\epsilon_B,
\boldsymbol\epsilon_C,\boldsymbol\epsilon_D)^T$
is the corresponding eight-component polarization vector. The signs in
$\boldsymbol{\mathcal J}_{\mathrm{AF}}$ are the sublattice signs
$(+1,-1,-1,+1)$ of Eq.~\eqref{eq:eta_cell}. The elastic dynamical matrix is a
$4\times4$ matrix of $2\times2$ blocks,
\begin{equation}
 \mathbf{D}(\mathbf k)=
 \begin{pmatrix}
 \mathbf{D}_{AA}&\mathbf{D}_{AB}&\mathbf{D}_{AC}&\mathbf{D}_{AD}\\
 \mathbf{D}_{AB}^{\dagger}&\mathbf{D}_{BB}&\mathbf{D}_{BC}&\mathbf{D}_{BD}\\
 \mathbf{D}_{AC}^{\dagger}&\mathbf{D}_{BC}^{\dagger}&\mathbf{D}_{CC}&\mathbf{D}_{CD}\\
 \mathbf{D}_{AD}^{\dagger}&\mathbf{D}_{BD}^{\dagger}&\mathbf{D}_{CD}^{\dagger}&\mathbf{D}_{DD}
 \end{pmatrix}.
 \label{eq:Dblock}
\end{equation}
Its diagonal blocks are
\begin{align}
 \mathbf{D}_{AA}=\mathbf{D}_{DD}
 &=\frac{1}{m}
 \begin{pmatrix}
 2k_1+k_2+k_3&k_2-k_3\\
 k_2-k_3&2k_1+k_2+k_3
 \end{pmatrix},
 \label{eq:DAA}\\
 \mathbf{D}_{BB}=\mathbf{D}_{CC}
 &=\frac{1}{m}
 \begin{pmatrix}
 2k_1+k_2+k_3&k_3-k_2\\
 k_3-k_2&2k_1+k_2+k_3
 \end{pmatrix}.
 \label{eq:DBB}
\end{align}
The axial off-diagonal blocks are
\begin{align}
 \mathbf{D}_{AB}=\mathbf{D}_{CD}
 &=-\frac{2k_1\cos(k_xa)}{m}
 \begin{pmatrix}1&0\\0&0\end{pmatrix},
 \label{eq:DAB}\\
 \mathbf{D}_{AC}=\mathbf{D}_{BD}
 &=-\frac{2k_1\cos(k_ya)}{m}
 \begin{pmatrix}0&0\\0&1\end{pmatrix}.
 \label{eq:DAC}
\end{align}
Finally, the diagonal-bond blocks are
\begin{align}
 \mathbf{D}_{AD}=-\frac{1}{m}
 \begin{pmatrix}
 k_2\cos[(k_x+k_y)a]+k_3\cos[(k_x-k_y)a]
 &k_2\cos[(k_x+k_y)a]-k_3\cos[(k_x-k_y)a]\\
 k_2\cos[(k_x+k_y)a]-k_3\cos[(k_x-k_y)a]
 &k_2\cos[(k_x+k_y)a]+k_3\cos[(k_x-k_y)a]
 \end{pmatrix},
 \label{eq:DAD}\\
 \mathbf{D}_{BC}=-\frac{1}{m}
 \begin{pmatrix}
 k_2\cos[(k_x-k_y)a]+k_3\cos[(k_x+k_y)a]
 &-k_2\cos[(k_x-k_y)a]+k_3\cos[(k_x+k_y)a]\\
 -k_2\cos[(k_x-k_y)a]+k_3\cos[(k_x+k_y)a]
 &k_2\cos[(k_x-k_y)a]+k_3\cos[(k_x+k_y)a]
 \end{pmatrix}.
 \label{eq:DBC}
\end{align}
Eqs.~\eqref{eq:qep}--\eqref{eq:DBC} specify the eight phonon branches in the reduced Brillouin zone $|k_x|,|k_y|<\pi/(2a)$. For arbitrary $k_2\neq k_3$ and $\beta\neq0$, the frequencies are the positive roots of
\begin{equation}
 \det\!\left[\mathbf{D}(\mathbf k)-\omega^2\mathbf{I}_8
 -2\ii\omega\beta\boldsymbol{\mathcal J}_{\mathrm{AF}}\right]=0.
 \label{eq:dispersion_determinant}
\end{equation}

\subsection{Closed-form dispersion of the uniform parent lattice}

The uniform parent lattice is obtained by setting $k_2=k_3=\bar k$ and removing the antiferromagnetic gyroscopic coupling, $\beta=0$. Its primitive cell contains one site, and the two in-plane polarizations are governed by the $2\times2$ dynamical matrix
\begin{equation}
 \mathbf{D}_0(\mathbf k)=\frac{2}{m}
 \begin{pmatrix}
 k_1[1-\cos(k_xa)]+\bar k[1-\cos(k_xa)\cos(k_ya)]
 &\bar k\sin(k_xa)\sin(k_ya)\\
 \bar k\sin(k_xa)\sin(k_ya)
 &k_1[1-\cos(k_ya)]+\bar k[1-\cos(k_xa)\cos(k_ya)]
 \end{pmatrix}.
 \label{eq:Dparent}
\end{equation}
Writing its three independent matrix elements as $D_{0,xx}$, $D_{0,yy}$, and $D_{0,xy}$, the two frequencies are
\begin{equation}
 \omega_{\pm}^2(\mathbf k)=
 \frac{D_{0,xx}+D_{0,yy}}{2}
 \pm\sqrt{\left(\frac{D_{0,xx}-D_{0,yy}}{2}\right)^2+D_{0,xy}^2}.
 \label{eq:parent_dispersion}
\end{equation}
The eight branches in the $2a\times2a$ representation are obtained by folding $\omega_\pm$ with the reciprocal shifts $(0,0)$, $(\pi/a,0)$, $(0,\pi/a)$, and $(\pi/a,\pi/a)$.

The off-diagonal element
\begin{equation}
 D_{0,xy}(\mathbf k)=\frac{2\bar k}{m}\sin(k_xa)\sin(k_ya)
 \label{eq:parent_mixing}
\end{equation}
mixes the two Cartesian polarizations. It vanishes when all diagonal springs are removed, in which case the $x$ and $y$ vibrations form independent harmonic sectors.

The two real polarization vectors can be written as
\begin{equation}
 \mathbf e_+(\mathbf k)=
 \begin{pmatrix}\cos\theta_{\mathbf k}\\ \sin\theta_{\mathbf k}\end{pmatrix},
 \qquad
 \mathbf e_-(\mathbf k)=
 \begin{pmatrix}-\sin\theta_{\mathbf k}\\ \cos\theta_{\mathbf k}\end{pmatrix},
 \label{eq:parent_eigenvectors}
\end{equation}
where
\begin{equation}
 \tan(2\theta_{\mathbf k})=
 \frac{2D_{0,xy}(\mathbf k)}{D_{0,xx}(\mathbf k)-D_{0,yy}(\mathbf k)}
 =\frac{2\bar k\sin(k_xa)\sin(k_ya)}
 {k_1[\cos(k_ya)-\cos(k_xa)]}.
 \label{eq:parent_polarization_angle}
\end{equation}
The diagonal springs therefore rotate the normal-mode polarization away from the Cartesian axes whenever both $k_x$ and $k_y$ are nonzero.

At long wavelength, Eq.~\eqref{eq:Dparent} becomes
\begin{equation}
 \mathbf{D}_0(\mathbf k)=\frac{a^2}{m}
 \begin{pmatrix}
 (k_1+\bar k)k_x^2+\bar k k_y^2&2\bar k k_xk_y\\
 2\bar k k_xk_y&\bar k k_x^2+(k_1+\bar k)k_y^2
 \end{pmatrix}+O(k^4a^4).
 \label{eq:longwave_D}
\end{equation}
Writing $\mathbf k=k(\cos\phi,\sin\phi)$ and substituting into Eq.~\eqref{eq:longwave_D}, the acoustic eigenvalues take the form $\omega_\pm^2=c_\pm^2(\phi)k^2$, which gives
\begin{equation}
 c_{\pm}^2(\phi)=\frac{a^2}{m}\left[
 \frac{k_1+2\bar k}{2}
 \pm\sqrt{\frac{k_1^2}{4}\cos^2(2\phi)+\bar k^2\sin^2(2\phi)}
 \right].
 \label{eq:sound_velocities}
\end{equation}
These velocities quantify the directional propagation and polarization mixing present in the uniform elastic parent lattice. Along the square axes they reduce to the longitudinal and transverse sound velocities, while away from the axes the diagonal springs rotate the principal polarizations. This establishes that anisotropic propagation and polarization mixing can occur even at $\beta=\delta=0$; the $d_{xy}$ splitter response identified below requires the additional staggered gyroscopic and checkerboard orders.

\subsection{Normalization, group velocity, and mode angular momentum}

Because Eq.~\eqref{eq:qep} is a quadratic eigenvalue problem, its natural mode norm differs from the Euclidean norm. Define
\begin{equation}
 \boldsymbol{\mathcal Q}(\omega,\mathbf k)
 =
 \mathbf{D}(\mathbf k)-\omega^2\mathbf{I}_8
 -2\ii\omega\beta\boldsymbol{\mathcal J}_{\mathrm{AF}} .
 \label{eq:qep_operator}
\end{equation}
For a real positive-frequency eigenmode, $\boldsymbol{\mathcal Q}$ is Hermitian. The homogeneous eigenvalue equation fixes the polarization only up to an overall complex scale. To determine that scale, consider the classical normal mode
\begin{equation}
 \mathbf u(t)=A\boldsymbol\epsilon_{n\mathbf k}e^{-\ii\omega_{n\mathbf k}t}
 +A^*\boldsymbol\epsilon_{n\mathbf k}^*e^{\ii\omega_{n\mathbf k}t}.
\end{equation}
Using $\boldsymbol{\mathcal Q}(\omega_{n\mathbf k},\mathbf k)\boldsymbol\epsilon_{n\mathbf k}=0$, its cycle-averaged energy is
\begin{align}
 \overline E
 &=\frac{m}{2}\overline{|\dot{\mathbf u}|^2}
 +\frac{m}{2}\overline{\mathbf u^T\mathbf{D}\mathbf u}\nonumber\\
 &=2m\omega_{n\mathbf k}^2|A|^2
 \boldsymbol\epsilon_{n\mathbf k}^{\dagger}
 \left[
 \mathbf{I}_8+\frac{\ii\beta}{\omega_{n\mathbf k}}\boldsymbol{\mathcal J}_{\mathrm{AF}}
 \right]
 \boldsymbol\epsilon_{n\mathbf k}.
 \label{eq:mode_energy}
\end{align}
The physical displacement is unchanged under $\boldsymbol\epsilon_{n\mathbf k}\to c\boldsymbol\epsilon_{n\mathbf k}$ and $A\to A/c$. We therefore choose the eigenvector scale such that
\begin{equation}
 \boldsymbol\epsilon_{n\mathbf k}^{\dagger}
 \left[
 \mathbf{I}_8+\frac{\ii\beta}{\omega_{n\mathbf k}}\boldsymbol{\mathcal J}_{\mathrm{AF}}
 \right]
 \boldsymbol\epsilon_{n\mathbf k}=1.
 \label{eq:norm}
\end{equation}
With this convention, $A$ is the mode amplitude such that Eq.~\eqref{eq:mode_energy} becomes $\overline E=2m\omega_{n\mathbf k}^2|A|^2$. Eq.~\eqref{eq:norm} is equivalently written as
\begin{equation}
 -\frac{1}{2\omega_{n\mathbf k}}\,
 \boldsymbol\epsilon_{n\mathbf k}^{\dagger}
 \left.
 \frac{\partial\boldsymbol{\mathcal Q}}{\partial\omega}
 \right|_{\omega_{n\mathbf k}}
 \boldsymbol\epsilon_{n\mathbf k}
 =1,
 \label{eq:qep_norm_general}
\end{equation}
because $\partial\boldsymbol{\mathcal Q}/\partial\omega=-2\omega\mathbf I_8-2\ii\beta\boldsymbol{\mathcal J}_{\mathrm{AF}}$. With the quantized expansion
\begin{equation}
 \hat{\mathbf u}(t)=
 \sqrt{\frac{\hbar}{2m\omega_{n\mathbf k}}}
 \left(
 \boldsymbol\epsilon_{n\mathbf k}\hat a_{n\mathbf k}e^{-\ii\omega_{n\mathbf k}t}
 +\boldsymbol\epsilon_{n\mathbf k}^*\hat a_{n\mathbf k}^{\dagger}e^{\ii\omega_{n\mathbf k}t}
 \right),
\end{equation}
Eq.~\eqref{eq:norm} gives $\hat H_{n\mathbf k}=\hbar\omega_{n\mathbf k}(\hat a_{n\mathbf k}^{\dagger}\hat a_{n\mathbf k}+1/2)$, so each phonon adds the energy $\hbar\omega_{n\mathbf k}$. At $\beta=0$, Eq.~\eqref{eq:norm} reduces to $\boldsymbol\epsilon_{n\mathbf k}^{\dagger}\boldsymbol\epsilon_{n\mathbf k}=1$.

To obtain the group velocity, differentiate
$\boldsymbol{\mathcal Q}(\omega_{n\mathbf k},\mathbf k)\boldsymbol\epsilon_{n\mathbf k}=0$
with respect to $k_a$ and left-multiply by
$\boldsymbol\epsilon_{n\mathbf k}^{\dagger}$. Hermiticity removes the derivative of the eigenvector and gives
\begin{equation}
 0=
 \boldsymbol\epsilon_{n\mathbf k}^{\dagger}
 \left[
 \frac{\partial\mathbf{D}}{\partial k_a}
 -2v_{n\mathbf k,a}
 \left(
 \omega_{n\mathbf k}\mathbf{I}_8+\ii\beta\boldsymbol{\mathcal J}_{\mathrm{AF}}
 \right)
 \right]
 \boldsymbol\epsilon_{n\mathbf k}.
 \label{eq:velocity_derivation}
\end{equation}
Using Eq.~\eqref{eq:norm}, we obtain
\begin{equation}
 v_{n\mathbf k,a}=\frac{1}{2\omega_{n\mathbf k}}
 \boldsymbol\epsilon_{n\mathbf k}^{\dagger}
 \frac{\partial\mathbf{D}(\mathbf k)}{\partial k_a}
 \boldsymbol\epsilon_{n\mathbf k},
 \qquad a=x,y.
 \label{eq:velocity}
\end{equation}

We define the uniform four-site rotation generator
$\boldsymbol{\mathcal J}=\operatorname{diag}(\mathbf{J},\mathbf{J},\mathbf{J},\mathbf{J})$.
The mechanical angular momentum carried by a normalized phonon mode is
\begin{equation}
 L_{n\mathbf k}^z
 =\hbar\boldsymbol\epsilon_{n\mathbf k}^{\dagger}
 \ii\boldsymbol{\mathcal J}
 \boldsymbol\epsilon_{n\mathbf k}.
 \label{eq:mode_L}
\end{equation}
At $\beta=0$, a nondegenerate elastic eigenvector can be chosen real. Since $\mathbf{J}$ is antisymmetric, a real vector $\mathbf x$ satisfies $\mathbf x^T\mathbf{J}\mathbf x=0$, and Eq.~\eqref{eq:mode_L} therefore vanishes.

\subsection{Angular momentum generated by staggered gyroscopic order}

Let $|n\rangle$ be a real eigenvector of the elastic checkerboard at $\beta=0$, with eigenfrequency $\omega_n$,
\begin{equation}
 \mathbf{D}(\mathbf k,\delta)|n\rangle=\omega_n^2|n\rangle.
 \label{eq:elastic_modes}
\end{equation}
Differentiating Eq.~\eqref{eq:qep} with respect to $\beta$ and projecting onto an elastic mode $|m\rangle$ gives
\begin{equation}
 \left.\frac{\partial|n\rangle}{\partial\beta}\right|_{\beta=0}
 =2\ii\omega_n\sum_{m\ne n}
 \frac{|m\rangle\langle m|\boldsymbol{\mathcal J}_{\mathrm{AF}}|n\rangle}
 {\omega_m^2-\omega_n^2}.
 \label{eq:beta_state}
\end{equation}
Substitution into Eq.~\eqref{eq:mode_L} gives
\begin{equation}
 L_{n\mathbf k}^z=
 4\hbar\beta\omega_n\sum_{m\ne n}
 \frac{
 \langle m|\boldsymbol{\mathcal J}_{\mathrm{AF}}|n\rangle
 \langle n|\boldsymbol{\mathcal J}|m\rangle
 }
 {\omega_n^2-\omega_m^2}
 +O(\beta^3).
 \label{eq:L_beta}
\end{equation}
Here the elastic eigenvectors are real, so both matrix elements in the numerator are real and
$\langle n|\ii\boldsymbol{\mathcal J}|m\rangle
=\ii\langle n|\boldsymbol{\mathcal J}|m\rangle$.
Thus the explicit factor $\ii$ and the imaginary-part operation cancel, yielding the real expression above.
This shows that staggered gyroscopic order generates phonon angular momentum by mixing elastic modes with different linear polarizations. Degenerate modes are treated by diagonalizing the perturbation within the complete degenerate subspace before applying Eq.~\eqref{eq:L_beta} to states outside that subspace.

\subsection{Checkerboard contribution and $d_{xy}$ symmetry}

We expand the elastic dynamical matrix to first order in the checkerboard contrast,
\begin{equation}
 \mathbf{D}(\mathbf k,\delta)=\mathbf{D}(\mathbf k,0)
 +\delta\left.\frac{\partial\mathbf{D}}{\partial\delta}\right|_{\delta=0}
 +O(\delta^2).
 \label{eq:Ddelta}
\end{equation}
The nonzero independent blocks of the derivative, evaluated at $\delta=0$, are
\begin{align}
 \left.\frac{\partial\mathbf{D}_{AA}}{\partial\delta}\right|_{\delta=0}
 =\left.\frac{\partial\mathbf{D}_{DD}}{\partial\delta}\right|_{\delta=0}
 &=\frac{2\bar k}{m}\begin{pmatrix}0&-1\\-1&0\end{pmatrix},
 \label{eq:Ddelta_AA}\\
 \left.\frac{\partial\mathbf{D}_{BB}}{\partial\delta}\right|_{\delta=0}
 =\left.\frac{\partial\mathbf{D}_{CC}}{\partial\delta}\right|_{\delta=0}
 &=\frac{2\bar k}{m}\begin{pmatrix}0&1\\1&0\end{pmatrix},
 \label{eq:Ddelta_diag}\\
 \left.\frac{\partial\mathbf{D}_{AD}}{\partial\delta}\right|_{\delta=0}
 &=\frac{\bar k}{m}
 \begin{pmatrix}
 \cos[(k_x+k_y)a]-\cos[(k_x-k_y)a]
 &\cos[(k_x+k_y)a]+\cos[(k_x-k_y)a]\\
 \cos[(k_x+k_y)a]+\cos[(k_x-k_y)a]
 &\cos[(k_x+k_y)a]-\cos[(k_x-k_y)a]
 \end{pmatrix},
 \label{eq:Ddelta_AD}\\
 \left.\frac{\partial\mathbf{D}_{BC}}{\partial\delta}\right|_{\delta=0}
 &=\frac{\bar k}{m}
 \begin{pmatrix}
 \cos[(k_x-k_y)a]-\cos[(k_x+k_y)a]
 &-\cos[(k_x-k_y)a]-\cos[(k_x+k_y)a]\\
 -\cos[(k_x-k_y)a]-\cos[(k_x+k_y)a]
 &\cos[(k_x-k_y)a]-\cos[(k_x+k_y)a]
 \end{pmatrix}.
 \label{eq:Ddelta_off}
\end{align}
For a nondegenerate elastic mode of the uniform parent lattice, the first-order frequency shift and polarization correction produced by the elastic checkerboard contrast are
\begin{align}
 \left.\frac{\partial\omega_n}{\partial\delta}\right|_{\delta=0}
 &=\frac{1}{2\omega_n}
 \left\langle n\left|
 \left.\frac{\partial\mathbf{D}}{\partial\delta}\right|_{\delta=0}
 \right|n\right\rangle,
 \label{eq:omega_delta}\\
 \left.\frac{\partial|n\rangle}{\partial\delta}\right|_{\delta=0}
 &=\sum_{p\ne n}|p\rangle
 \frac{\left\langle p\left|
 \left.\frac{\partial\mathbf{D}}{\partial\delta}\right|_{\delta=0}
 \right|n\right\rangle}
 {\omega_n^2-\omega_p^2}.
 \label{eq:state_delta}
\end{align}
Eqs.~\eqref{eq:omega_delta} and \eqref{eq:state_delta} apply to isolated positive-frequency modes. At acoustic zero modes and band degeneracies, the perturbation is first diagonalized within the complete degenerate subspace, and the derivatives are evaluated for the resulting branches.

Differentiating Eq.~\eqref{eq:L_beta} with respect to $\delta$ gives the leading angular momentum that requires both the gyroscopic order and the elastic checkerboard contrast,
\begin{equation}
 L_{n\mathbf k,\mathrm{cb}}^z
 =\hbar\beta\delta\,\Lambda_n(\mathbf k)
 +O(\beta^3\delta,\beta\delta^3),
 \label{eq:L_bd}
\end{equation}
where
\begin{equation}
 \Lambda_n(\mathbf k)=
 \left.\frac{\partial}{\partial\delta}\left
 \{4\omega_n\sum_{m\ne n}
 \frac{
 \langle m|\boldsymbol{\mathcal J}_{\mathrm{AF}}|n\rangle
 \langle n|\boldsymbol{\mathcal J}|m\rangle
 }
 {\omega_n^2-\omega_m^2}\right\}\right|_{\delta=0}.
 \label{eq:Lambda}
\end{equation}
All frequencies and eigenvectors inside the derivative are those of the elastic problem at $\beta=0$. The derivative in Eq.~\eqref{eq:Lambda} is expanded into parent-lattice band sums in the appendix section on the checkerboard angular-momentum coefficient.

The same contribution can be isolated for any observable $R$ by combining the four antiferromagnetic and elastic domains,
\begin{equation}
 R_{\mathrm{cb}}(\beta,\delta)=\frac14\left[
 R(\beta,\delta)-R(-\beta,\delta)
 -R(\beta,-\delta)+R(-\beta,-\delta)\right].
 \label{eq:domain_projection}
\end{equation}
By construction, $R_{\mathrm{cb}}$ is odd under reversal of either order and even under simultaneous reversal.

For an isolated nondegenerate branch, the checkerboard-projected angular momentum obeys
\begin{align}
 L_{n,\mathrm{cb}}^z(-k_x,k_y)&=-L_{n,\mathrm{cb}}^z(k_x,k_y),&
 L_{n,\mathrm{cb}}^z(k_x,-k_y)&=-L_{n,\mathrm{cb}}^z(k_x,k_y),\\
 L_{n,\mathrm{cb}}^z(k_y,k_x)&=L_{n,\mathrm{cb}}^z(k_x,k_y),&
 L_{n,\mathrm{cb}}^z(-k_y,k_x)&=-L_{n,\mathrm{cb}}^z(k_x,k_y).
 \label{eq:dxy_relations}
\end{align}
The first symmetry-allowed harmonic in the reduced Brillouin zone can therefore be written as
\begin{equation}
 \Lambda_n(\mathbf k)=F_n(\mathbf k)\sin(2k_xa)\sin(2k_ya),
 \qquad
 L_{n\mathbf k,\mathrm{cb}}^z=
 \hbar\beta\delta\,F_n(\mathbf k)\sin(2k_xa)\sin(2k_ya)
 +O(\beta^3\delta,\beta\delta^3),
 \label{eq:dxy_form}
\end{equation}
where $F_n(\mathbf k)$ is even under $k_x\to-k_x$, $k_y\to-k_y$, and $k_x\leftrightarrow k_y$. In the notation of the square-lattice point group, this transformation law belongs to the one-dimensional $B_2$ representation, whose lowest-order basis function is $k_xk_y$ and which is conventionally identified with $d_{xy}$ symmetry. Higher harmonics with the same $B_2$ symmetry are contained in the momentum dependence of $F_n$. Eq.~\eqref{eq:dxy_form} therefore has nodes on the $k_x$ and $k_y$ axes and opposite signs in adjacent quadrants. At a degeneracy, the same idea applies to the symmetry-adapted combinations obtained by diagonalizing within the degenerate subspace.

\section{Local angular-momentum balance and bond currents}
\label{sec:continuity}

\subsection{Local balance and canonical density}

The canonical momentum in Eq.~\eqref{eq:EL_derivatives} is
$\mathbf p_i=m\dot{\mathbf u}_i+m\beta_i\mathbf{J}\mathbf u_i$. The angular momentum generated by rotations of the canonical pair $(\mathbf u_i,\mathbf p_i)$ is therefore
\begin{equation}
 \lambda_i^z=\mathbf u_i\times\mathbf p_i
 =L_i^z+m\beta_i|\mathbf u_i|^2,
 \qquad
 \mathbf a\times\mathbf b\equiv a_xb_y-a_yb_x.
 \label{eq:lambda}
\end{equation}
Thus $\lambda_i^z$ contains the mechanical angular momentum $L_i^z$ and the gyroscopic contribution $m\beta_i|\mathbf u_i|^2$.

Taking the cross product of $\mathbf u_i$ with Eq.~\eqref{eq:eom_site} gives
\begin{equation}
 \frac{\dd L_i^z}{\dd t}
 +m\beta_i\frac{\dd}{\dd t}|\mathbf u_i|^2
 +\sum_{j\in\mathcal N_i}
 \mathbf u_i\times\boldsymbol{\Phi}_{ij}(\mathbf u_i-\mathbf u_j)
 =
 \tau_i^{\mathrm{bath}},
 \label{eq:crossed_eom}
\end{equation}
where we used
\begin{equation}
 \mathbf u_i\times
 \left(2m\beta_i\mathbf{J}\dot{\mathbf u}_i\right)
 =
 2m\beta_i\,\mathbf u_i\cdot\dot{\mathbf u}_i
 =
 m\beta_i\frac{\dd}{\dd t}|\mathbf u_i|^2,
 \label{eq:gyro_derivative}
\end{equation}
and defined the torque exerted by the reservoir on site $i$ as
\begin{equation}
 \tau_i^{\mathrm{bath}}
 =
 \mathbf u_i\times
 \left(\boldsymbol\xi_i-m\gamma_i\dot{\mathbf u}_i\right).
 \label{eq:bath_torque}
\end{equation}
The first two terms of Eq.~\eqref{eq:crossed_eom} are exactly
$\dd\lambda_i^z/\dd t$.

\subsection{Exact transfer and torque associated with one bond}

For a bond $i-j$, define
$\Delta\mathbf u_{ij}=\mathbf u_i-\mathbf u_j$ and
$\mathbf f_{ij}=\boldsymbol{\Phi}_{ij}\Delta\mathbf u_{ij}$.
The spring force on $i$ is $-\mathbf f_{ij}$ and the force on $j$ is
$+\mathbf f_{ij}$. Their torques about the origin are therefore
\begin{equation}
 \tau_{i\leftarrow j}^L=-\mathbf u_i\times\mathbf f_{ij},
 \qquad
 \tau_{j\leftarrow i}^L=\mathbf u_j\times\mathbf f_{ij}.
 \label{eq:endpoint_torques}
\end{equation}
We separate these two torques into their antisymmetric and symmetric combinations,
\begin{align}
 j_{i\to j}^L
 &=
 \frac{\tau_{j\leftarrow i}^L-\tau_{i\leftarrow j}^L}{2}
 =
 \frac12(\mathbf u_i+\mathbf u_j)\times
 \boldsymbol{\Phi}_{ij}(\mathbf u_i-\mathbf u_j),
 \label{eq:bond_current}\\
 q_{ij}^L
 &=
 \frac{\tau_{i\leftarrow j}^L+\tau_{j\leftarrow i}^L}{2}
 =
 -\frac12(\mathbf u_i-\mathbf u_j)\times
 \boldsymbol{\Phi}_{ij}(\mathbf u_i-\mathbf u_j).
 \label{eq:bond_torque}
\end{align}
The first quantity is a bond current because
$j_{j\to i}^L=-j_{i\to j}^L$: it transfers angular momentum between the two sites and cancels when both endpoints are included. The second quantity satisfies $q_{ji}^L=q_{ij}^L$ and measures the net torque exerted by the anisotropic spring on the two displacement degrees of freedom. Explicitly,
\begin{equation}
 \tau_{i\leftarrow j}^L=q_{ij}^L-j_{i\to j}^L,
 \qquad
 \tau_{j\leftarrow i}^L=q_{ij}^L+j_{i\to j}^L,
 \qquad
 \tau_{i\leftarrow j}^L+\tau_{j\leftarrow i}^L=2q_{ij}^L.
 \label{eq:bond_torque_decomposition}
\end{equation}
The static lattice frame that fixes the bond direction experiences the opposite reaction torque, $-2q_{ij}^L$. A central spring conserves angular momentum within the two moving endpoints only when $q_{ij}^L=0$.

Using Eq.~\eqref{eq:bond_torque_decomposition} in Eq.~\eqref{eq:crossed_eom} gives the local balance equation
\begin{equation}
 \frac{\dd\lambda_i^z}{\dd t}
 +\sum_{j\in\mathcal N_i}j_{i\to j}^L
 =
 \sum_{j\in\mathcal N_i}q_{ij}^L
 +\tau_i^{\mathrm{bath}}.
 \label{eq:local_balance}
\end{equation}
The left-hand side contains the local canonical density and the angular momentum transferred to neighboring sites. The right-hand side contains the torque transferred between the moving lattice and the static elastic frame, together with the torque exchanged with the reservoirs.

For an axial $x$ bond of stiffness $k_1$, Eqs.~\eqref{eq:bond_current} and \eqref{eq:bond_torque} give
\begin{align}
 j_{i\to j}^{L,x\text{-bond}}
 &=-\frac{k_1}{2}(u_{iy}+u_{jy})(u_{ix}-u_{jx}),\\
 q_{ij}^{L,x\text{-bond}}
 &=\frac{k_1}{2}(u_{ix}-u_{jx})(u_{iy}-u_{jy}).
 \label{eq:x_bond}
\end{align}
For an axial $y$ bond,
\begin{align}
 j_{i\to j}^{L,y\text{-bond}}
 &=\frac{k_1}{2}(u_{ix}+u_{jx})(u_{iy}-u_{jy}),\\
 q_{ij}^{L,y\text{-bond}}
 &=-\frac{k_1}{2}(u_{ix}-u_{jx})(u_{iy}-u_{jy}).
 \label{eq:y_bond}
\end{align}
For a diagonal $(1,1)$ bond of stiffness $k$,
\begin{align}
 j_{i\to j}^{L,+}
 &=\frac{k}{4}\big[(u_{ix}+u_{jx})-(u_{iy}+u_{jy})\big]
 \big[(u_{ix}-u_{jx})+(u_{iy}-u_{jy})\big],\\
 q_{ij}^{L,+}
 &=-\frac{k}{4}\big[(u_{ix}-u_{jx})^2-(u_{iy}-u_{jy})^2\big].
 \label{eq:plus_bond}
\end{align}
For a diagonal $(1,-1)$ bond,
\begin{align}
 j_{i\to j}^{L,-}
 &=-\frac{k}{4}\big[(u_{ix}+u_{jx})+(u_{iy}+u_{jy})\big]
 \big[(u_{ix}-u_{jx})-(u_{iy}-u_{jy})\big],\\
 q_{ij}^{L,-}
 &=\frac{k}{4}\big[(u_{ix}-u_{jx})^2-(u_{iy}-u_{jy})^2\big].
 \label{eq:minus_bond}
\end{align}
The diagonal spring constant in Eqs.~\eqref{eq:plus_bond} and \eqref{eq:minus_bond} is $k_2$ or $k_3$ according to the plaquette of the checkerboard structure.

\subsection{Global and first-moment balances}

Summing Eq.~\eqref{eq:local_balance} over all sites gives
\begin{align}
 \sum_i\sum_{j\in\mathcal N_i}j_{i\to j}^L
 &=\sum_{\langle ij\rangle}
 \left(j_{i\to j}^L+j_{j\to i}^L\right)=0,\\
 \sum_i\sum_{j\in\mathcal N_i}q_{ij}^L
 &=2\sum_{\langle ij\rangle}q_{ij}^L.
 \label{eq:pair_sums}
\end{align}
Hence
\begin{equation}
 \frac{\dd}{\dd t}\sum_i\lambda_i^z
 =
 2\sum_{\langle ij\rangle}q_{ij}^L
 +\sum_i\tau_i^{\mathrm{bath}}.
 \label{eq:global_balance}
\end{equation}
The internal transfer cancels pairwise, whereas the bond torque appears twice because it acts with the same sign at the two endpoints, as shown in Eq.~\eqref{eq:bond_torque_decomposition}.

In a steady state, multiplying Eq.~\eqref{eq:local_balance} by $\mathbf r_i$ and summing over sites gives
\begin{equation}
 \sum_{\langle ij\rangle}(\mathbf r_j-\mathbf r_i)j_{i\to j}^L
 +\sum_i\mathbf r_i
 \left(
 \sum_{j\in\mathcal N_i}q_{ij}^L+\tau_i^{\mathrm{bath}}
 \right)=0.
 \label{eq:first_moment}
\end{equation}
Eq.~\eqref{eq:first_moment} gives the steady-state balance between angular momentum transferred along bonds and torques exchanged with the elastic frame and thermal reservoirs.

\section{Thermal transport}
\label{sec:exact_transport}

\subsection{Exact harmonic response}

For a finite checkerboard containing $N_{\mathrm s}$ sites, we collect the $2N_{\mathrm s}$ displacement components into $\mathbf u$. Eq.~\eqref{eq:eom_site} becomes
\begin{equation}
 \mathbf{M}\ddot{\mathbf u}
 +2\mathbf{M}\mathbf{B}\dot{\mathbf u}
 +\mathbf{M}\boldsymbol{\Gamma}\dot{\mathbf u}
 +\mathbf{K}\mathbf u
 =\boldsymbol\xi,
 \label{eq:matrix_eom}
\end{equation}
where
\begin{equation}
 \mathbf{M}=m\mathbf{I}_{2N_{\mathrm s}},\qquad
 \mathbf{B}=\operatorname{diag}(\beta_1\mathbf{J},\ldots,\beta_{N_{\mathrm s}}\mathbf{J}),\qquad
 \boldsymbol{\Gamma}=\operatorname{diag}(\gamma_1\mathbf{I}_2,\ldots,\gamma_{N_{\mathrm s}}\mathbf{I}_2).
 \label{eq:finite_matrices}
\end{equation}
Here $\mathbf{M}$ is the mass matrix, $\mathbf{B}$ contains the local gyroscopic rates, and $\boldsymbol{\Gamma}$ is the damping-rate matrix. We also define the uniform $N_{\mathrm s}$-site rotation generator
$\boldsymbol{\mathcal J}_{N_{\mathrm s}}=\mathbf{I}_{N_{\mathrm s}}\otimes\mathbf{J}=\operatorname{diag}(\mathbf{J},\ldots,\mathbf{J})$.
The force-constant matrix is assembled from the bond tensors as
\begin{equation}
 \mathbf{K}_{ii}=\sum_{j\in\mathcal N_i}\boldsymbol{\Phi}_{ij},
 \qquad
 \mathbf{K}_{ij}=-\boldsymbol{\Phi}_{ij}
 \quad\text{for a connected pair }i-j.
 \label{eq:K_blocks}
\end{equation}
The bond tensors in Eqs.~\eqref{eq:Phi_xy} and \eqref{eq:Phi_diag}, together with the complete bond list in Table~\ref{tab:bonds}, determine every block of $\mathbf{K}$ in terms of $k_1$, $k_2$, and $k_3$.

A free finite lattice has two zero-frequency modes corresponding to uniform translations in the $x$ and $y$ directions. A stationary displacement covariance therefore requires these rigid translations to be removed. This can be done by fixing the center of mass,
$\sum_i\mathbf u_i=0$, by projecting onto relative coordinates with
\begin{equation}
 \mathbf{P}=
 \mathbf{I}_{2N_{\mathrm s}}
 -\frac{1}{N_{\mathrm s}}
 \left(\mathbf 1_{N_{\mathrm s}}\mathbf 1_{N_{\mathrm s}}^T\right)\otimes\mathbf{I}_2,
 \qquad
 \mathbf u\rightarrow\mathbf{P}\mathbf u,
 \label{eq:translation_projector}
\end{equation}
or by adding an arbitrarily weak pinning potential. All inverses and stationary covariances below are evaluated on this stable subspace.

Using the Fourier convention
$\mathbf u(t)=\int\dd\omega\,\mathbf u(\omega)e^{-\ii\omega t}/(2\pi)$,
Eq.~\eqref{eq:matrix_eom} gives
\begin{equation}
 \mathbf u(\omega)=\mathbf{G}(\omega)\boldsymbol\xi(\omega),
 \qquad
 \mathbf{G}(\omega)=
 \left[
 \mathbf{K}-\omega^2\mathbf{M}
 -2\ii\omega\mathbf{M}\mathbf{B}
 -\ii\omega\mathbf{M}\boldsymbol{\Gamma}
 \right]^{-1}.
 \label{eq:Green}
\end{equation}
We use $\mathbf{G}$ throughout to denote the retarded displacement Green function. Its $2\times2$ block $\mathbf{G}_{is}(\omega)$ gives the displacement at site $i$ generated by a force applied at reservoir site $s$.

For independent Ohmic reservoirs, the symmetrized force-noise spectrum at site $s$ is
\begin{equation}
 \frac12\left\langle
 \xi_{s\mu}(\omega)\xi_{s\nu}^{\dagger}(\omega')
 +\xi_{s\nu}^{\dagger}(\omega')\xi_{s\mu}(\omega)
 \right\rangle
 =
 2\pi\delta(\omega-\omega')
 m\gamma_s\hbar\omega
 \coth\!\left(\frac{\hbar\omega}{2k_{\mathrm B}T_s}\right)
 \delta_{\mu\nu}.
 \label{eq:quantum_noise}
\end{equation}
For $k_{\mathrm B}T_s\gg\hbar|\omega|$, the factor
$\hbar\omega\coth[\hbar\omega/(2k_{\mathrm B}T_s)]$
approaches $2k_{\mathrm B}T_s$, and Eq.~\eqref{eq:quantum_noise} reduces to the classical white-noise correlator in Eq.~\eqref{eq:classical_noise}.

We define the temperature derivative of the noise spectrum at the background temperature $T_0$ as
\begin{equation}
 \chi_s(\omega,T_0)
 \equiv
 \left.
 \frac{\partial}{\partial T}
 \left[
 m\gamma_s\hbar\omega
 \coth\!\left(\frac{\hbar\omega}{2k_{\mathrm B}T}\right)
 \right]\right|_{T_0}
 =
 \frac{m\gamma_s(\hbar\omega)^2}{2k_{\mathrm B}T_0^2}
 \operatorname{csch}^2\!\left(
 \frac{\hbar\omega}{2k_{\mathrm B}T_0}
 \right).
 \label{eq:noise_derivative}
\end{equation}
Its classical limit is $\chi_s\to2m\gamma_sk_{\mathrm B}$.

For weak reservoir-temperature changes
$T_s=T_0+\Delta T_s$, the induced displacement-covariance response is
\begin{equation}
 \Delta\langle\mathbf u_i\mathbf u_j^T\rangle
 =
 \sum_s\Delta T_s
 \int_{-\infty}^{\infty}\frac{\dd\omega}{2\pi}\,
 \chi_s(\omega,T_0)
 \mathbf{G}_{is}(\omega)\mathbf{G}_{js}^{\dagger}(\omega).
 \label{eq:Cuu_arbitrary_profile}
\end{equation}
This equation applies to any prescribed reservoir-temperature profile, including the one-sided and center-hot profiles used for the finite-sample accumulation maps.

For a weak linear temperature profile,
\begin{equation}
 T_s=T_0+(\mathbf r_s-\mathbf r_0)\cdot\boldsymbol\nabla T.
 \label{eq:Tprofile}
\end{equation}
Choosing $\mathbf r_0$ as their mean position makes the imposed temperature changes sum to zero and separates the gradient response from a uniform temperature shift. The covariance induced per unit gradient along direction $b$ is
\begin{equation}
 \mathbf{C}_{uu,ij}^{(b)}
 \equiv
 \left.
 \frac{\partial\langle\mathbf u_i\mathbf u_j^T\rangle}
 {\partial(\partial_bT)}
 \right|_{\boldsymbol\nabla T=0}
 =
 \sum_s(r_{s,b}-r_{0,b})
 \int_{-\infty}^{\infty}\frac{\dd\omega}{2\pi}\,
 \chi_s(\omega,T_0)
 \mathbf{G}_{is}(\omega)\mathbf{G}_{js}^{\dagger}(\omega).
 \label{eq:Cuu_exact}
\end{equation}
This expression contains the complete harmonic response, including intermode correlations and all boundary effects.

The corresponding displacement--velocity block is
\begin{equation}
 \mathbf{C}_{uv,ij}^{(b)}
 =
 \sum_s(r_{s,b}-r_{0,b})
 \int_{-\infty}^{\infty}\frac{\dd\omega}{2\pi}\,
 \ii\omega\,\chi_s(\omega,T_0)
 \mathbf{G}_{is}(\omega)\mathbf{G}_{js}^{\dagger}(\omega),
 \label{eq:Cuv_exact}
\end{equation}
and the local mechanical PAM response is
\begin{equation}
 L_i^{z,(b)}
 \equiv
 \frac{\partial\langle L_i^z\rangle}{\partial(\partial_bT)}
 =
 m\left[
 \left(\mathbf{C}_{uv,ii}^{(b)}\right)_{xy}
 -
 \left(\mathbf{C}_{uv,ii}^{(b)}\right)_{yx}
 \right].
 \label{eq:Li_cov}
\end{equation}

The area-averaged PAM current density is obtained by weighting each bond current by its bond displacement,
\begin{equation}
 J_a^L=
 \frac{1}{\mathcal A}
 \sum_{\langle ij\rangle}
 (r_{j,a}-r_{i,a})
 \langle j_{i\to j}^L\rangle,
 \qquad
 J_a^L=-\sum_b\sigma_{ab}^L\partial_bT.
 \label{eq:J_sigma}
\end{equation}
Here $\mathcal A$ is the sample area, $a$ labels the direction of angular-momentum flow, $b$ labels the direction of the temperature gradient, and $\sigma_{ab}^L=-\partial J_a^L/\partial(\partial_bT)$ defines the PAM conductivity.

Using Eq.~\eqref{eq:bond_current} and substituting Eq.~\eqref{eq:Cuu_exact}, the exact bond-transfer conductivity is
\begin{align}
 \sigma_{ab}^L
 &=
 \frac{1}{2\mathcal A}
 \sum_{\langle ij\rangle}
 (r_{j,a}-r_{i,a})
 \sum_s(r_{s,b}-r_{0,b})
 \int_{-\infty}^{\infty}\frac{\dd\omega}{2\pi}\,
 \chi_s(\omega,T_0)
 \nonumber\\[-2mm]
 &\qquad\times
 \operatorname{Re}\Tr\!\left[
 \mathbf{J}\boldsymbol{\Phi}_{ij}
 \big(\mathbf{G}_{is}-\mathbf{G}_{js}\big)
 \big(\mathbf{G}_{is}+\mathbf{G}_{js}\big)^{\dagger}
 \right].
 \label{eq:sigma_exact}
\end{align}
The elastic parameters enter explicitly through $\boldsymbol{\Phi}_{ij}$ and through the force-constant matrix $\mathbf{K}$ contained in $\mathbf{G}$. The staggered gyroscopic order enters through $\mathbf{B}$ in the same Green function.

The gradient response of the bond torque is
\begin{align}
 \frac{\partial\langle q_{ij}^L\rangle}{\partial(\partial_bT)}
 &=
 \frac12
 \sum_s(r_{s,b}-r_{0,b})
 \int_{-\infty}^{\infty}\frac{\dd\omega}{2\pi}\,
 \chi_s(\omega,T_0)
 \nonumber\\[-2mm]
 &\qquad\times
 \operatorname{Re}\Tr\!\left[
 \mathbf{J}\boldsymbol{\Phi}_{ij}
 \big(\mathbf{G}_{is}-\mathbf{G}_{js}\big)
 \big(\mathbf{G}_{is}-\mathbf{G}_{js}\big)^{\dagger}
 \right].
 \label{eq:q_exact}
\end{align}
Together with the reservoir-torque response, these quantities obey the gradient derivative of Eq.~\eqref{eq:first_moment}. In the classical additive white-noise convention, the equal-time position--noise correlator vanishes and
\begin{equation}
 \frac{\partial\langle\tau_i^{\mathrm{bath}}\rangle}
 {\partial(\partial_bT)}
 =
 -\gamma_i L_i^{z,(b)}.
 \label{eq:bath_torque_response}
\end{equation}
For quantum reservoirs, the corresponding symmetrized position--noise term is retained explicitly in the absorbed edge-torque expression below.

\subsection{Energy balance and edge response}

To complete the transport description, we also derive the accompanying energy balance. Assign half of each bond energy to each endpoint and define
\begin{equation}
 \varepsilon_i=
 \frac{m}{2}|\dot{\mathbf u}_i|^2
 +\frac14\sum_{j\in\mathcal N_i}
 (\mathbf u_i-\mathbf u_j)^T
 \boldsymbol{\Phi}_{ij}
 (\mathbf u_i-\mathbf u_j).
 \label{eq:local_energy}
\end{equation}
Taking the scalar product of Eq.~\eqref{eq:eom_site} with
$\dot{\mathbf u}_i$ gives
\begin{equation}
 \frac{\dd}{\dd t}\langle\varepsilon_i\rangle
 +\sum_{j\in\mathcal N_i}\langle j_{i\to j}^{E}\rangle
 =
 P_i^{\mathrm{bath}},
 \label{eq:energy_balance}
\end{equation}
where
\begin{align}
 j_{i\to j}^{E}
 &=
 \frac12(\dot{\mathbf u}_i+\dot{\mathbf u}_j)^T
 \boldsymbol{\Phi}_{ij}(\mathbf u_i-\mathbf u_j),
 \qquad
 j_{j\to i}^{E}=-j_{i\to j}^{E},
 \label{eq:energy_bond_current}\\
 P_i^{\mathrm{bath}}
 &=
 \left\langle\dot{\mathbf u}_i\cdot\boldsymbol\xi_i\right\rangle
 -m\gamma_i\left\langle|\dot{\mathbf u}_i|^2\right\rangle.
 \label{eq:bath_power}
\end{align}
Here $j_{i\to j}^{E}$ is the energy current leaving site $i$ through the bond $i-j$, and $P_i^{\mathrm{bath}}$ is the net power delivered by the reservoir at site $i$ to the lattice. The term $\langle\dot{\mathbf u}_i\cdot\boldsymbol\xi_i\rangle$ is the stochastic power injected by the fluctuating force, whereas $m\gamma_i\langle|\dot{\mathbf u}_i|^2\rangle$ is the power removed by viscous damping. Hence $P_i^{\mathrm{bath}}>0$ means that reservoir $i$ injects energy into the lattice.

The gyroscopic force performs no work because
$\dot{\mathbf u}_i^T\mathbf{J}\dot{\mathbf u}_i=0$.
In a stationary state, summing Eq.~\eqref{eq:energy_balance} over all sites gives
$\sum_iP_i^{\mathrm{bath}}=0$, which expresses global energy conservation.

The torque absorbed at edge $\alpha$ is the negative of the reservoir torque in Eq.~\eqref{eq:bath_torque}, summed with the edge-selection weights $w_{\alpha i}$:
\begin{equation}
 \tau_{\alpha}^{\mathrm{abs}}
 =
 \sum_i w_{\alpha i}
 \left[
 \gamma_i\langle L_i^z\rangle
 -\frac12\left\langle
 \mathbf u_i\times\boldsymbol\xi_i
 -\boldsymbol\xi_i\times\mathbf u_i
 \right\rangle
 \right].
 \label{eq:terminal_torque_thermal}
\end{equation}
The symmetrized second term is required when displacement and force are quantum operators. For the classical additive white-noise dynamics used in the finite-sample calculations,
$\langle\mathbf u_i\times\boldsymbol\xi_i\rangle=0$, and therefore
\begin{equation}
 -\frac{\partial\tau_{\alpha}^{\mathrm{abs}}}
 {\partial(\partial_bT)}
 =
 -\sum_i w_{\alpha i}\gamma_i L_i^{z,(b)}.
 \label{eq:terminal_G}
\end{equation}
To compare opposite boundaries, we define
\begin{equation}
 \mathcal T_x=
 \frac{\tau_{\mathrm{right}}^{\mathrm{abs}}
 -\tau_{\mathrm{left}}^{\mathrm{abs}}}{2\ell_y},
 \qquad
 \mathcal T_y=
 \frac{\tau_{\mathrm{top}}^{\mathrm{abs}}
 -\tau_{\mathrm{bottom}}^{\mathrm{abs}}}{2\ell_x},
 \qquad
 G_{ab}^L=
 -\frac{\partial\mathcal T_a}{\partial(\partial_bT)}.
 \label{eq:terminal_tensor}
\end{equation}
Here $\ell_x$ and $\ell_y$ are the sample dimensions. The edge-response tensor $\mathbf{G}^L$ describes the absorbed-torque response at opposite edges and is the finite-sample counterpart of the bond conductivity $\boldsymbol{\sigma}^L$. The two tensors have the same symmetry decomposition but need not be numerically identical.

\subsection{Classical covariance and modal decomposition}

In the classical white-noise limit, all equal-time displacement and velocity correlations can be obtained from a stationary covariance matrix. With
$\mathbf z=(\mathbf u,\dot{\mathbf u})^T$,
\begin{equation}
 \dot{\mathbf z}=\mathbf{A}\mathbf z+\boldsymbol\zeta,
 \qquad
 \mathbf{A}=
 \begin{pmatrix}
 0&\mathbf{I}_{2N_{\mathrm s}}\\
 -\mathbf{M}^{-1}\mathbf{K}&-2\mathbf{B}-\boldsymbol{\Gamma}
 \end{pmatrix},
 \qquad
 \boldsymbol\zeta=
 \begin{pmatrix}0\\ \mathbf{M}^{-1}\boldsymbol\xi\end{pmatrix}.
 \label{eq:A_matrix}
\end{equation}
The stationary covariance
$\mathbf{C}=\langle\mathbf z\mathbf z^T\rangle$
obeys
$\mathbf{A}\mathbf{C}+\mathbf{C}\mathbf{A}^T=-\mathbf{Q}$, where
$\mathbf{Q}=\langle\boldsymbol\zeta(t)\boldsymbol\zeta(t')^T\rangle/\delta(t-t')$.
Differentiating this equation with respect to a temperature gradient gives
\begin{equation}
 \mathbf{A}\mathbf{C}^{(b)}
 +\mathbf{C}^{(b)}\mathbf{A}^T
 =-\mathbf{Q}^{(b)},
 \qquad
 \mathbf{C}^{(b)}
 \equiv
 \frac{\partial\langle\mathbf z\mathbf z^T\rangle}
 {\partial(\partial_bT)}.
 \label{eq:Lyapunov}
\end{equation}
Only the velocity--velocity block of the noise derivative is nonzero,
\begin{equation}
 \left(\mathbf{Q}_{vv}^{(b)}\right)_{i\mu,j\nu}
 =
 \frac{2\gamma_i k_{\mathrm B}(r_{i,b}-r_{0,b})}{m}
 \delta_{ij}\delta_{\mu\nu}.
 \label{eq:Qb}
\end{equation}
After the translational zero modes have been removed or weak pinning has been added, stability requires every eigenvalue of $\mathbf{A}$ to have a negative real part. The Lyapunov equation then has a unique solution. Formally,
\begin{equation}
 \operatorname{vec}\mathbf{C}^{(b)}
 =
 -\left[
 \mathbf{I}\otimes\mathbf{A}+\mathbf{A}\otimes\mathbf{I}
 \right]^{-1}
 \operatorname{vec}\mathbf{Q}^{(b)}.
 \label{eq:Lyapunov_solution}
\end{equation}
Here $\operatorname{vec}$ stacks the columns of a matrix into a vector and $\otimes$ denotes the Kronecker product. In practice, Eq.~\eqref{eq:Lyapunov} is solved directly without constructing the Kronecker-product inverse. Substitution of the displacement block of $\mathbf{C}^{(b)}$ into Eqs.~\eqref{eq:bond_current} and \eqref{eq:bond_torque} reproduces the classical limit of Eqs.~\eqref{eq:sigma_exact} and \eqref{eq:q_exact}; its displacement--velocity block gives Eq.~\eqref{eq:Li_cov}.

The separation between changes in mode populations and correlations between different modes can be evaluated analytically for the non-gyroscopic reference lattice with uniform damping $\gamma$. Let $\mathbf U_{\mu}$ be its mass-normalized normal modes,
\begin{equation}
 \mathbf{K}\mathbf U_{\mu}=\Omega_{\mu}^2\mathbf{M}\mathbf U_{\mu},
 \qquad
 \mathbf U_{\mu}^T\mathbf{M}\mathbf U_{\nu}=\delta_{\mu\nu},
 \qquad
 \mathbf u=\sum_{\mu}Q_{\mu}\mathbf U_{\mu},
 \label{eq:finite_normal_modes}
\end{equation}
where $Q_\mu$ is the modal coordinate and $\mathbf U_{i\mu}$ denotes the two Cartesian components of mode $\mu$ at site $i$. A temperature gradient produces the modal noise matrix
\begin{equation}
 W_{\mu\nu}^{(b)}
 =
 2\gamma k_{\mathrm B}\sum_i
 m(r_{i,b}-r_{0,b})\,
 \mathbf U_{i\mu}\cdot\mathbf U_{i\nu}.
 \label{eq:modal_noise_matrix}
\end{equation}
For $\Delta_{\mu\nu}=\Omega_{\mu}^2-\Omega_{\nu}^2$, the covariance derivatives are
\begin{align}
 \langle Q_{\mu}Q_{\nu}\rangle^{(b)}
 &=
 \frac{2\gamma W_{\mu\nu}^{(b)}}
 {\Delta_{\mu\nu}^2+2\gamma^2(\Omega_{\mu}^2+\Omega_{\nu}^2)},
 \label{eq:modal_qq}\\
 \langle Q_{\mu}\dot Q_{\nu}\rangle^{(b)}
 &=
 \frac{\Delta_{\mu\nu}W_{\mu\nu}^{(b)}}
 {\Delta_{\mu\nu}^2+2\gamma^2(\Omega_{\mu}^2+\Omega_{\nu}^2)},
 \label{eq:modal_qv}\\
 \langle\dot Q_{\mu}\dot Q_{\nu}\rangle^{(b)}
 &=
 \frac{\gamma(\Omega_{\mu}^2+\Omega_{\nu}^2)W_{\mu\nu}^{(b)}}
 {\Delta_{\mu\nu}^2+2\gamma^2(\Omega_{\mu}^2+\Omega_{\nu}^2)}.
 \label{eq:modal_vv}
\end{align}
The diagonal terms $\mu=\nu$ describe changes in modal occupation. Eq.~\eqref{eq:modal_qv} vanishes on the diagonal, so in this real eigenmode basis the thermally induced angular momentum is carried by correlations between distinct modes. This diagonal--off-diagonal separation depends on the chosen modal basis; a contribution diagonal in the gyroscopic eigenmode basis can appear off diagonal in the real modes of the non-gyroscopic reference lattice.

\subsection{Checkerboard splitter tensor}

Applying the domain projection in Eq.~\eqref{eq:domain_projection} to the exact conductivity in Eq.~\eqref{eq:sigma_exact} isolates the response that is odd in both $\beta$ and $\delta$. The momentum-space result in Eq.~\eqref{eq:dxy_form} shows that this channel transforms according to the $B_2$, or $d_{xy}$, representation of the square lattice. Let
\begin{equation}
 \mathbf{M}_x=
 \begin{pmatrix}-1&0\\0&1\end{pmatrix},
 \qquad
 \mathbf{M}_d=
 \begin{pmatrix}0&1\\1&0\end{pmatrix}
 \label{eq:B2_operations}
\end{equation}
represent reflection about the $y$ axis and interchange of the two coordinate axes. A $B_2$ tensor is odd under the first operation and even under the second,
\begin{equation}
 \mathbf{M}_x\boldsymbol{\sigma}_{\mathrm{cb}}^L\mathbf{M}_x^T
 =-\boldsymbol{\sigma}_{\mathrm{cb}}^L,
 \qquad
 \mathbf{M}_d\boldsymbol{\sigma}_{\mathrm{cb}}^L\mathbf{M}_d^T
 =\boldsymbol{\sigma}_{\mathrm{cb}}^L.
 \label{eq:B2_tensor_constraints}
\end{equation}
Starting from a general $2\times2$ tensor, the first relation sets both diagonal components to zero, and the second makes the two off-diagonal components equal. Therefore
\begin{equation}
 \boldsymbol{\sigma}_{\mathrm{cb}}^L=
 D\begin{pmatrix}0&1\\1&0\end{pmatrix},
 \qquad
 D=\frac{\sigma_{xy}^{L,\mathrm{cb}}+\sigma_{yx}^{L,\mathrm{cb}}}{2}.
 \label{eq:D_tensor}
\end{equation}
A finite sample can also have response components outside this symmetry channel. The coefficient $D$ isolates the odd-in-$\beta$, odd-in-$\delta$ symmetric off-diagonal contribution.

In terms of the four domains, the exact coefficient is
\begin{equation}
 D=\frac18\sum_{s_\beta=\pm1}\sum_{s_\delta=\pm1}
 s_\beta s_\delta\left[
 \sigma_{xy}^L(s_\beta\beta,s_\delta\delta)
 +\sigma_{yx}^L(s_\beta\beta,s_\delta\delta)\right],
 \label{eq:D_domains}
\end{equation}
where each conductivity on the right-hand side is given by Eq.~\eqref{eq:sigma_exact} with
$k_2=\bar k(1-s_\delta\delta)$,
$k_3=\bar k(1+s_\delta\delta)$, and
$\beta_i=s_\beta\beta\eta_i$.

For a temperature gradient directed from hot to cold at an angle $\phi$ from the $x$ axis,
\begin{equation}
 -\boldsymbol\nabla T
 =
 |\boldsymbol\nabla T|
 \begin{pmatrix}\cos\phi\\ \sin\phi\end{pmatrix}.
 \label{eq:angle_gradient}
\end{equation}
Using $\mathbf J^L=-\boldsymbol{\sigma}_{\mathrm{cb}}^L\boldsymbol\nabla T$ gives
\begin{equation}
 \mathbf J^L=
 D|\boldsymbol\nabla T|
 \begin{pmatrix}\sin\phi\\ \cos\phi\end{pmatrix}.
 \label{eq:D_current_vector}
\end{equation}
Projecting onto
$\hat{\mathbf e}_{\parallel}=(\cos\phi,\sin\phi)^T$
and
$\hat{\mathbf e}_{\perp}=(-\sin\phi,\cos\phi)^T$
then yields
\begin{equation}
 J_{\parallel}^L
 =
 \hat{\mathbf e}_{\parallel}\cdot\mathbf J^L
 =
 D|\boldsymbol\nabla T|\sin(2\phi),
 \qquad
 J_{\perp}^L
 =
 \hat{\mathbf e}_{\perp}\cdot\mathbf J^L
 =
 D|\boldsymbol\nabla T|\cos(2\phi).
 \label{eq:angular_response}
\end{equation}
A gradient along $x$ or $y$ therefore produces a transverse splitter current, while a diagonal gradient produces a longitudinal PAM-Seebeck response. In the diagonal axes
\begin{equation}
 x'=\frac{x+y}{\sqrt2},\qquad
 y'=\frac{-x+y}{\sqrt2},
 \label{eq:rot_axes}
\end{equation}
the same tensor is
\begin{equation}
 \boldsymbol{\sigma}_{\mathrm{cb}}^{L\prime}=
 \begin{pmatrix}D&0\\0&-D\end{pmatrix}.
 \label{eq:D_rot}
\end{equation}
Thus the two diagonal directions carry opposite longitudinal PAM conductivities.

The domain definition gives
\begin{equation}
 D(0,\delta)=0,\qquad D(\beta,0)=0,
 \label{eq:D_nulls}
\end{equation}
and
\begin{equation}
 D(-\beta,\delta)=-D(\beta,\delta),\qquad
 D(\beta,-\delta)=-D(\beta,\delta),\qquad
 D(-\beta,-\delta)=D(\beta,\delta).
 \label{eq:D_reversal}
\end{equation}

\subsection{Convective mode-population limit}

The exact Green-function expression calculates the bond current without assuming independent phonon populations. A complementary weak-scattering approximation describes the PAM carried by modes moving with their group velocities. Because the local balance also includes bond torques, this convective contribution need not equal the bond current in Eq.~\eqref{eq:J_sigma}. They agree when intermode coherence, bond-torque contributions, and boundary effects can be neglected.

The linearized Boltzmann equation gives
\begin{equation}
 \delta n_{n\mathbf k}=-\tau_{n\mathbf k}
 \mathbf v_{n\mathbf k}\cdot\boldsymbol\nabla T
 \frac{\partial n_{\mathrm B}(\omega_{n\mathbf k},T)}{\partial T},
 \qquad
 \frac{\partial n_{\mathrm B}}{\partial T}
 =\frac{\hbar\omega_{n\mathbf k}}{k_{\mathrm B}T^2}
 n_{\mathrm B}(n_{\mathrm B}+1).
 \label{eq:delta_n}
\end{equation}
The resulting convective PAM current is
\begin{equation}
 J_a^{L,\mathrm{mode}}=
 \sum_n\int_{\mathrm{RBZ}}\frac{\dd^2k}{(2\pi)^2}
 L_{n\mathbf k}^zv_{n\mathbf k,a}\,\delta n_{n\mathbf k},
 \label{eq:J_mode}
\end{equation}
and the corresponding mode-population conductivity is
\begin{equation}
 \sigma_{ab}^{L,\mathrm{mode}}(T)=
 \sum_n\int_{\mathrm{RBZ}}\frac{\dd^2k}{(2\pi)^2}
 \tau_{n\mathbf k}L_{n\mathbf k}^z
 v_{n\mathbf k,a}v_{n\mathbf k,b}
 \frac{\partial n_{\mathrm B}}{\partial T}.
 \label{eq:sigma_mode}
\end{equation}
This tensor is symmetric because every term contains $v_av_b$.  The two domain reversals map corresponding phonon frequencies and group velocities into one another.  Assuming that the relaxation times respect the same domain symmetries, applying the domain projection and using Eq.~\eqref{eq:L_bd} gives
\begin{equation}
 D_{\mathrm{mode}}(T)=
 \sum_n\int_{\mathrm{RBZ}}\frac{\dd^2k}{(2\pi)^2}
 \tau_{n\mathbf k}L_{n\mathbf k,\mathrm{cb}}^z
 v_{n\mathbf k,x}v_{n\mathbf k,y}
 \frac{\partial n_{\mathrm B}}{\partial T}.
 \label{eq:D_mode}
\end{equation}
At weak gyroscopic order and weak checkerboard contrast,
\begin{equation}
 D_{\mathrm{mode}}(T)=\beta\delta\hbar
 \sum_n\int_{\mathrm{RBZ}}\frac{\dd^2k}{(2\pi)^2}
 \tau_{n\mathbf k}^{(0)}\Lambda_n(\mathbf k)
 v_{n\mathbf k,x}^{(0)}v_{n\mathbf k,y}^{(0)}
 \frac{\partial n_{\mathrm B}(\omega_{n\mathbf k}^{(0)},T)}{\partial T}
 +O(\beta^3\delta,\beta\delta^3).
 \label{eq:D_weak}
\end{equation}
The superscript $(0)$ denotes the uniform elastic parent lattice.  Eq.~\eqref{eq:D_weak} is the convective mode-population contribution associated with the equilibrium $d_{xy}$ angular-momentum texture.  The exact splitter coefficient is obtained from Eq.~\eqref{eq:D_domains}; in a finite thermal system it also contains the effects of the microscopic bond-current vertex, lattice torque, intermode correlations, and boundaries.

\section{Polarization-resolved coherent response}
\label{sec:coherent}

A monochromatic force provides a second way to probe the checkerboard splitter.  We write the force and displacement as
\begin{equation}
 \mathbf f(t)=\operatorname{Re}\!\left[\mathbf f(\omega)e^{-\ii\omega t}\right],
 \qquad
 \mathbf u(t)=\operatorname{Re}\!\left[\mathbf u(\omega)e^{-\ii\omega t}\right].
 \label{eq:harmonic_fields}
\end{equation}
The equation of motion gives
\begin{equation}
 \mathbf u(\omega)=\mathbf{G}(\omega)\mathbf f(\omega),
 \label{eq:coherent_solution}
\end{equation}
with the same Green function as Eq.~\eqref{eq:Green}.  The poles of $\mathbf{G}$ are the damped checkerboard phonon resonances, so the frequency dependence of every coherent observable follows from the spectrum derived above.

We describe an arbitrary polarization of a two-component source by
\begin{equation}
 \mathbf e(\vartheta,\varphi)=
 \begin{pmatrix}\cos\vartheta\\ e^{\ii\varphi}\sin\vartheta\end{pmatrix},
 \qquad \mathbf e^{\dagger}\mathbf e=1.
 \label{eq:source_polarization}
\end{equation}
A source distributed over one or more sites is represented by a $2N_{\mathrm s}\times2$ matrix $\mathbf{S}$, so that $\mathbf f(\omega)=f_0\mathbf{S}\mathbf e$.  Linear $x$ and $y$ excitation correspond to $\vartheta=0$ and $\vartheta=\pi/2$.  The circular polarizations are $\vartheta=\pi/4$, $\varphi=\pm\pi/2$.

For the complex displacement amplitude at site $i$, the time-averaged mechanical angular momentum is
\begin{equation}
 \overline{L_i^z}=m\omega\operatorname{Im}(u_{ix}^{*}u_{iy})
 =\frac{m\omega}{2}\mathbf u_i^{\dagger}(\ii\mathbf{J})\mathbf u_i.
 \label{eq:coherent_local_pam}
\end{equation}
If $\mathbf{P}_i$ selects the two displacement components at site $i$, the polarization-resolved local PAM is
\begin{equation}
 \overline{L_i^z}=\mathbf e^{\dagger}\mathbf{R}_i^{L}(\omega)\mathbf e,
 \qquad
 \mathbf{R}_i^{L}(\omega)=\frac{m|f_0|^2\omega}{2}\mathbf{S}^{\dagger}
 \mathbf{G}(\omega)^{\dagger}\mathbf{P}_i^T(\ii\mathbf{J})\mathbf{P}_i
 \mathbf{G}(\omega)\mathbf{S}.
 \label{eq:local_pam_response_matrix}
\end{equation}
The polarization dependence derived below applies to both the local PAM and the edge response.

Let $\boldsymbol{\Gamma}_{\alpha}$ denote the damping-rate matrix weighted by the edge-selection factors, with the $i$th site block $w_{\alpha i}\gamma_i\mathbf I_2$. The power and angular momentum absorbed at edge $\alpha$ are
\begin{align}
 P_{\alpha}(\omega,\mathbf e)
 &=\frac{\omega^2}{2}\mathbf u^{\dagger}\mathbf{M}\boldsymbol{\Gamma}_{\alpha}\mathbf u,
 \label{eq:terminal_power_coherent}\\
 \tau_{\alpha}^{\mathrm{abs}}(\omega,\mathbf e)
 &=\frac{\omega}{2}\mathbf u^{\dagger}
 \mathbf{M}\boldsymbol{\Gamma}_{\alpha}(\ii\boldsymbol{\mathcal J}_{N_{\mathrm s}})\mathbf u.
 \label{eq:terminal_torque_coherent}
\end{align}
Because $\mathbf{M}=m\mathbf I_{2N_{\mathrm s}}$ and the damping is isotropic within each site, $\mathbf{M}\boldsymbol{\Gamma}_{\alpha}$ commutes with $\boldsymbol{\mathcal J}_{N_{\mathrm s}}$, and both expressions are real.  The injected power is
\begin{equation}
 P_{\mathrm{in}}=\frac{\omega}{2}\operatorname{Im}\!\left[
 \mathbf f^{\dagger}\mathbf u\right].
 \label{eq:injected_power}
\end{equation}
In a steady harmonic state, $P_{\mathrm{in}}$ equals the power absorbed by all damped sites.

We characterize the circular polarization absorbed at an edge by the dimensionless ratio
\begin{equation}
 \Pi_{\alpha}(\omega,\mathbf e)=
 \frac{\omega\tau_{\alpha}^{\mathrm{abs}}(\omega,\mathbf e)}
 {P_{\alpha}(\omega,\mathbf e)}.
 \label{eq:terminal_polarization}
\end{equation}
For a locally circular displacement, $\Pi_{\alpha}=+1$ or $-1$, while a linearly polarized displacement gives zero.

Substituting Eq.~\eqref{eq:coherent_solution} into Eqs.~\eqref{eq:terminal_power_coherent} and \eqref{eq:terminal_torque_coherent} gives two $2\times2$ Hermitian response matrices,
\begin{align}
 \mathbf{R}_{\alpha}^{P}(\omega)
 &=\frac{|f_0|^2\omega^2}{2}\mathbf{S}^{\dagger}
 \mathbf{G}(\omega)^{\dagger}\mathbf{M}\boldsymbol{\Gamma}_{\alpha}
 \mathbf{G}(\omega)\mathbf{S},
 \label{eq:power_response_matrix}\\
 \mathbf{R}_{\alpha}^{\tau}(\omega)
 &=\frac{|f_0|^2\omega}{2}\mathbf{S}^{\dagger}
 \mathbf{G}(\omega)^{\dagger}\mathbf{M}\boldsymbol{\Gamma}_{\alpha}
 (\ii\boldsymbol{\mathcal J}_{N_{\mathrm s}})\mathbf{G}(\omega)\mathbf{S}.
 \label{eq:torque_response_matrix}
\end{align}
They give
\begin{equation}
 P_{\alpha}=\mathbf e^{\dagger}\mathbf{R}_{\alpha}^{P}\mathbf e,
 \qquad
 \tau_{\alpha}^{\mathrm{abs}}=\mathbf e^{\dagger}\mathbf{R}_{\alpha}^{\tau}\mathbf e.
 \label{eq:quadratic_polarization_response}
\end{equation}
Eqs.~\eqref{eq:power_response_matrix}--\eqref{eq:quadratic_polarization_response} give the complete polarization-resolved response of the checkerboard for an arbitrary source profile and frequency.

For any Hermitian $2\times2$ response matrix $\boldsymbol{\mathcal O}$, write
\begin{equation}
 \boldsymbol{\mathcal O}=o_0\mathbf{I}_2+o_1\boldsymbol{\sigma}_z+o_2\boldsymbol{\sigma}_x+o_3\boldsymbol{\sigma}_y,
 \label{eq:stokes_decomposition}
\end{equation}
where $o_0=\Tr\boldsymbol{\mathcal O}/2$, $o_1=(\boldsymbol{\mathcal O}_{xx}-\boldsymbol{\mathcal O}_{yy})/2$, $o_2=\operatorname{Re}\boldsymbol{\mathcal O}_{xy}$, and $o_3=-\operatorname{Im}\boldsymbol{\mathcal O}_{xy}$.  Eq.~\eqref{eq:source_polarization} then gives
\begin{equation}
 \mathbf e^{\dagger}\boldsymbol{\mathcal O}\mathbf e
 =o_0+o_1\cos(2\vartheta)
 +o_2\sin(2\vartheta)\cos\varphi
 +o_3\sin(2\vartheta)\sin\varphi.
 \label{eq:stokes_response}
\end{equation}
The $x$ and $y$ responses are $o_0\pm o_1$, the responses to linear polarization along the two diagonals are $o_0\pm o_2$, and the two circular responses are $o_0\pm o_3$.  An incoherent unpolarized source is the equal average of $x$ and $y$ excitation and gives $o_0$.

The checkerboard part of the coherent torque is isolated by applying the domain projection to the response matrix,
\begin{equation}
 \mathbf{R}_{\alpha}^{\tau,\mathrm{cb}}=
 \frac14\left[
 \mathbf{R}_{\alpha}^{\tau}(\beta,\delta)
 -\mathbf{R}_{\alpha}^{\tau}(-\beta,\delta)
 -\mathbf{R}_{\alpha}^{\tau}(\beta,-\delta)
 +\mathbf{R}_{\alpha}^{\tau}(-\beta,-\delta)\right].
 \label{eq:coherent_torque_projection}
\end{equation}
The positive normalization is obtained from the domain-even absorbed power,
\begin{equation}
 \mathbf{R}_{\alpha}^{P,\mathrm{even}}=
 \frac14\sum_{s_{\beta}=\pm1}\sum_{s_{\delta}=\pm1}
 \mathbf{R}_{\alpha}^{P}(s_{\beta}\beta,s_{\delta}\delta).
 \label{eq:coherent_power_even}
\end{equation}
The symmetry-isolated edge polarization is therefore
\begin{equation}
 \Pi_{\alpha}^{\mathrm{cb}}(\omega,\mathbf e)=
 \omega\frac{\mathbf e^{\dagger}\mathbf{R}_{\alpha}^{\tau,\mathrm{cb}}\mathbf e}
 {\mathbf e^{\dagger}\mathbf{R}_{\alpha}^{P,\mathrm{even}}\mathbf e}.
 \label{eq:coherent_projected_polarization}
\end{equation}
For the upper and lower edges, the antisymmetric splitter signal and the common polarization are
\begin{equation}
 \mathcal S(\omega,\mathbf e)=\frac{\Pi_{\mathrm{top}}^{\mathrm{cb}}-\Pi_{\mathrm{bottom}}^{\mathrm{cb}}}{2},
 \qquad
 \mathcal C(\omega,\mathbf e)=\frac{\Pi_{\mathrm{top}}^{\mathrm{cb}}+\Pi_{\mathrm{bottom}}^{\mathrm{cb}}}{2}.
 \label{eq:coherent_splitting}
\end{equation}
The quantities $\mathcal S$ and $\mathcal C$ measure the difference and common part of the two edge signals. Opposite-edge polarizations are not necessarily equal and opposite: this also depends on the excitation polarization, not only the device geometry. Reversing either the antiferromagnetic or checkerboard domain reverses $\mathcal S$, while simultaneous reversal restores its sign. Circular excitation can add a common helicity-dependent response, but the domain projection in Eq.~\eqref{eq:coherent_torque_projection} retains only the response that is odd under both domain reversals.

\section{PAM Hall response}
\label{sec:PAMHE}

The antisymmetric part of the exact conductivity is
\begin{equation}
 H=\frac{\sigma_{yx}^L-\sigma_{xy}^L}{2},
 \qquad
 \boldsymbol{\sigma}_{\mathrm{PAMHE}}^L=
 H\begin{pmatrix}0&-1\\1&0\end{pmatrix}.
 \label{eq:H}
\end{equation}
It is absent from Eq.~\eqref{eq:sigma_mode} because the convective mode-population tensor is symmetric.  The exact Green-function response is not restricted to the symmetric convective form and includes the intermode coherences that can generate an antisymmetric conductivity~\cite{ParkYang2020,Bustamante2026}.

When $k_2=k_3=0$, the elastic parent-lattice mixing element in Eq.~\eqref{eq:parent_mixing} vanishes.  At finite $\beta$, however, the staggered gyroscopic term still couples the two Cartesian polarizations.  The absence of diagonal springs does not by itself determine $H$.  Its value must be obtained from the exact response; its vanishing can follow from the symmetries of the bulk or device geometry or from a parameter-dependent cancellation.

Retaining the checkerboard splitter ($B_2$) and Hall ($A_2$) contributions, the two-channel part of the conductivity is
\begin{equation}
 \left.\boldsymbol{\sigma}^L\right|_{D+H}=
 \begin{pmatrix}0&D-H\\D+H&0\end{pmatrix}.
 \label{eq:full_sigma}
\end{equation}
The same symmetry analysis applies to the edge-response tensor $\mathbf{G}^L$ used in Fig.~2 of the main text.
For a thermal gradient at angle $\phi$, the combined longitudinal and transverse currents are
\begin{equation}
 J_{\parallel}^L=D|\boldsymbol\nabla T|\sin(2\phi),
 \qquad
 J_{\perp}^L=|\boldsymbol\nabla T|\left[H+D\cos(2\phi)\right].
 \label{eq:combined_angular_response}
\end{equation}
A diagonal gradient, $\phi=\pi/4$, separates the two coefficients: the longitudinal current measures $D$ and the transverse current measures $H$.

\section{Numerical evaluation}
\label{sec:numerical}

We evaluate the classical finite-sample response on an open square lattice with $N$ sites along each Cartesian direction and $N^2$ sites in total. For each antiferromagnetic and elastic domain $(s_\beta,s_\delta)$, we assemble the force-constant matrix $\mathbf K$, the gyroscopic-rate matrix $\mathbf B$, the damping-rate matrix $\boldsymbol\Gamma$, and the noise covariance associated with the imposed temperature profile. The state vector $\mathbf z=(\mathbf u,\dot{\mathbf u})^T$ has dimension $4N^2$, and the gradient-induced covariance response $\mathbf C^{(b)}$ is obtained by solving the continuous Lyapunov equation, Eq.~\eqref{eq:Lyapunov}, directly on the stable subspace.

We obtain the transport tensors $\sigma_{ab}^L$ and $G_{ab}^L$ from linear temperature gradients along $x$ and $y$. Smooth one-sided and center-hot temperature profiles give the spatial PAM accumulation and current maps. The covariance yields local PAM from its displacement--velocity block and bond currents and torques from its displacement block, using Eqs.~\eqref{eq:Li_cov}, \eqref{eq:bond_current}, and \eqref{eq:bond_torque}. We calculate the area-averaged bond current $J_a^L$, absorbed edge torque $\tau_{\alpha}^{\mathrm{abs}}$, and reservoir power $P_i^{\mathrm{bath}}$, and check the balances in Eqs.~\eqref{eq:first_moment} and \eqref{eq:energy_balance}. Calculating all four sign domains separates contributions according to their parity under $\beta\to-\beta$ and $\delta\to-\delta$. The splitter is the odd--odd $B_2$ contribution; the Hall component in the parameter scan and size-convergence analysis is the even--even $A_2$ contribution. The main-text real-space accumulation maps instead use the raw physical-domain response. For the supplementary current maps, the $d$-wave-like field uses the odd--odd bond current, while the Hall-like field uses the antisymmetric part of the physical $(+\beta,+\delta)$ response.

\begin{figure*}[t]
    \centering
    \includegraphics[width=\textwidth]{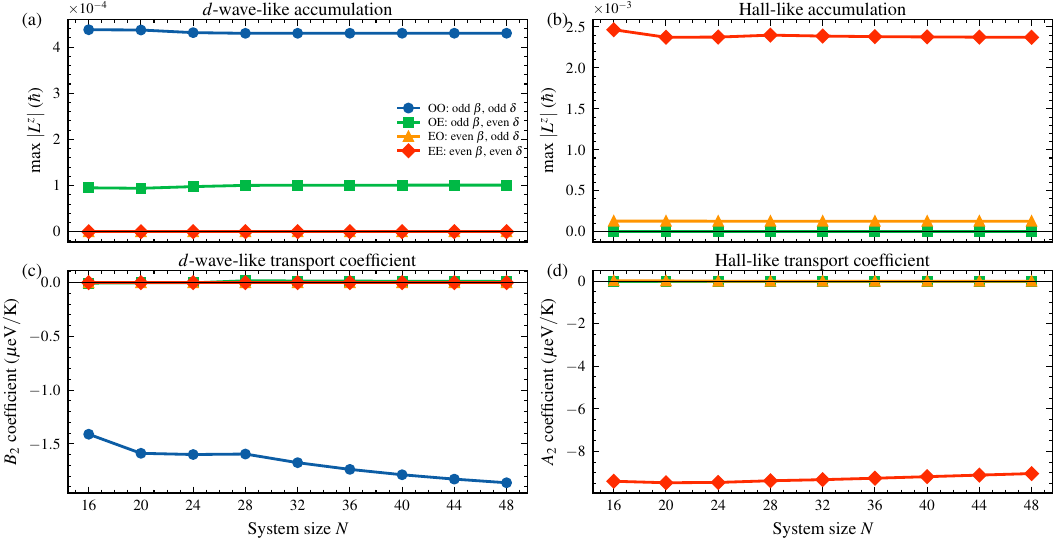}
    \caption{System-size dependence of the symmetry-resolved PAM response. (a,b) Maximum local magnitudes of the $d$-wave-like and Hall-like accumulations under a one-sided smooth temperature profile. (c,d) Corresponding $B_2$ (symmetric off-diagonal) and $A_2$ (antisymmetric off-diagonal) edge-response coefficients extracted from linear temperature gradients. The dominant splitter belongs to the odd-$\beta$, odd-$\delta$ (OO) channel, and the dominant Hall contribution to the even-$\beta$, even-$\delta$ (EE) channel. The dominant signals vary smoothly with system size. Smaller mixed-parity (OE and EO) accumulation components are visible, but their edge-response coefficients are negligible on the scale of the dominant channels, consistent with finite-size or edge-related contributions without demonstrating spatial localization.}
    \label{fig:system_size_convergence}
\end{figure*}

Fig.~\ref{fig:system_size_convergence} compares accumulation amplitudes from the one-sided temperature profile with edge-response coefficients from linear gradients. The dominant splitter and Hall contributions vary smoothly with sample size. Smaller mixed-parity contributions are visible in the local accumulation, but their edge-response coefficients are negligible compared with the dominant channels. This is consistent with finite-size or edge effects, without establishing their spatial localization.

\begin{figure*}[t]
    \centering
    \includegraphics[width=0.85\textwidth]{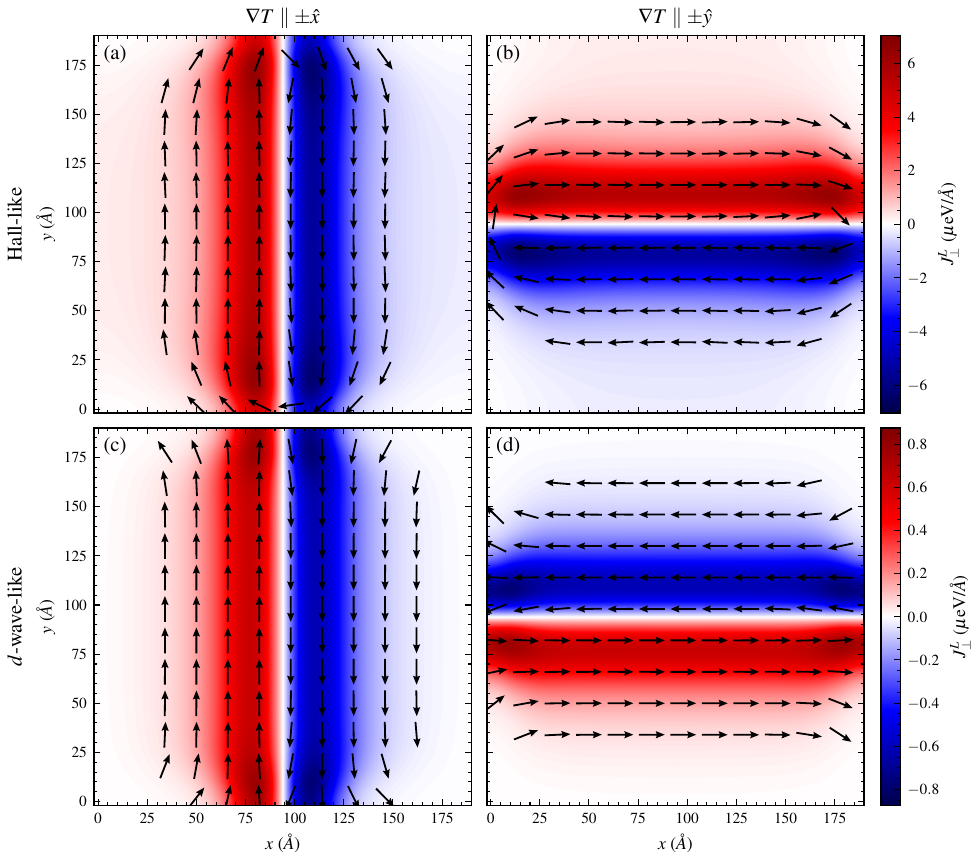}
    \caption{Numerically reconstructed local symmetry-resolved phonon-angular-momentum (PAM) current for the center-hot temperature profile in the $N=48$ sample. The left and right columns correspond to the two profile orientations, $\nabla T\parallel\pm\hat{x}$ and $\nabla T\parallel\pm\hat{y}$, respectively; the two signs indicate the oppositely directed local gradients on the two sides of the central hot region. The upper row shows the Hall-like $A_2$ contribution, whereas the lower row shows the $d$-wave-like $B_2$ contribution. The color scale shows the Cartesian current component perpendicular to the profile axis, $J_y^L$ for the $x$-oriented profile and $J_x^L$ for the $y$-oriented profile. The arrows show the direction of the full local PAM-current vector $(J_x^L,J_y^L)$; their lengths are normalized for visibility. For the $90^\circ$-rotated profile, the dominant $d$-wave-like transverse-current pattern is predominantly opposite to the Hall-like pattern over the central transport region, as seen from the interchange of the positive and negative lobes and the reversal of the corresponding current directions.}
    \label{fig:supp_local_pam_currents}
\end{figure*}

To visualize the microscopic current pattern, we convert the scalar current on each undirected bond into a site-centered vector density,
\begin{equation}
 \mathbf j_i^L=\frac{1}{2a^2}\sum_{j\in\mathcal N_i}
 (\mathbf r_j-\mathbf r_i)\langle j_{i\to j}^L\rangle .
 \label{eq:local_current_map}
\end{equation}
This symmetric endpoint assignment preserves the exact first moment because $a^2\sum_i\mathbf j_i^L=\sum_{\langle ij\rangle}(\mathbf r_j-\mathbf r_i)\langle j_{i\to j}^L\rangle$. Fig.~\ref{fig:supp_local_pam_currents} applies this construction to a center-hot profile in the $N=48$ sample. The two profile orientations generate local gradients along $\pm\hat x$ and $\pm\hat y$. The color map shows the component orthogonal to the profile axis and the arrows show the complete local vector direction. Under a $90^\circ$ rotation of the thermal profile, the dominant $d$-wave-like pattern is predominantly opposite to the Hall-like pattern in the central transport region, consistent with their symmetric $B_2$ and antisymmetric $A_2$ transformation laws.

\begin{table}[t]
\centering
\caption{Parameters used for the numerical calculations. The damping is increased smoothly from $\gamma_{\mathrm{bulk}}$ in the interior to the total edge value $\gamma_{\mathrm{edge}}$ at the outer boundary to enhance absorption of outgoing lattice excitations and their angular momentum and thereby suppress artificial reflections from the finite edges.}
\label{tab:numerical_parameters}
\begin{tabular}{@{}lll@{}}
\toprule
Quantity & Symbol & Value\\
\midrule
Atomic mass & $m$ & $28.0855\,\mathrm{amu}$\\
Lattice constant & $a$ & $4.0\,\text{\AA}$\\
Axial spring constant & $k_1$ & $4.0\,\mathrm{N\,m^{-1}}$\\
Mean diagonal spring constant & $\bar k$ & $2.0\,\mathrm{N\,m^{-1}}$\\
Checkerboard contrast & $\delta$ & $0.8$\\
Diagonal spring constants & $(k_2,k_3)$ & $(0.4,3.6)\,\mathrm{N\,m^{-1}}$\\
Gyroscopic frequency & $\beta/(2\pi)$ & $0.3\,\mathrm{THz}$\\
Bulk and edge damping rates & $(\gamma_{\mathrm{bulk}},\gamma_{\mathrm{edge}})$ & $(2.0,4.0)\,\mathrm{ps^{-1}}$\\
Pinning frequency & $\omega_{\mathrm{pin}}/(2\pi)$ & $0.005\,\mathrm{THz}$\\
Reference, hot, and cold temperatures & $(T_0,T_{\mathrm H},T_{\mathrm C})$ & $(300,310,300)\,\mathrm{K}$\\
\bottomrule
\end{tabular}
\end{table}

\begin{table}[t]
\centering
\small
\caption{Temperature profiles used in the numerical calculations. The linear profile is used to extract transport coefficients, the one-sided hot-to-cold profile is used for accumulation and parameter scans, and the center-hot profile is used for real-space accumulation and current maps.}
\label{tab:temperature_profiles}
\renewcommand{\arraystretch}{1.2}
\begin{tabular}{lll}
\toprule
\parbox[t]{0.12\textwidth}{Profile} &
\parbox[t]{0.39\textwidth}{Temperature field} &
\parbox[t]{0.21\textwidth}{Use and parameters} \\
\midrule

\parbox[t]{0.12\textwidth}{Linear gradient} &
\parbox[t]{0.39\textwidth}{Let \(s_i=\mathbf r_i\cdot\hat{\mathbf e}\) be the coordinate along the driving direction. We impose
\(\displaystyle \Delta T_i=-|\boldsymbol{\nabla}T|(s_i-\bar s_\gamma)\), where
\(\displaystyle \bar s_\gamma=\frac{\sum_j\gamma_j s_j}{\sum_j\gamma_j}\). Subtracting \(\bar s_\gamma\) removes the uniform-temperature component.} &
\parbox[t]{0.21\textwidth}{Used to obtain the transport tensors and the coefficients \(D\) and \(H\), with a unit gradient along \(x\) or \(y\).} \\[1ex]

\parbox[t]{0.12\textwidth}{One-sided hot-to-cold} &
\parbox[t]{0.39\textwidth}{Let \(s_{\mathrm H}\) and \(s_{\mathrm C}\) denote the hot and cold boundaries. Define
\(\displaystyle f(s)=\frac12\left[1-\tanh\!\left(\frac{s-s_0}{w}\right)\right]\)
and
\(\displaystyle p_{\mathrm{1s}}(s)=\frac{f(s)-f(s_{\mathrm C})}{f(s_{\mathrm H})-f(s_{\mathrm C})}\).
The profile is
\(\displaystyle T_{\mathrm{1s}}(s)=T_{\mathrm C}+(T_{\mathrm H}-T_{\mathrm C})p_{\mathrm{1s}}(s)\),
so that \(T_{\mathrm{1s}}(s_{\mathrm H})=T_{\mathrm H}\) and \(T_{\mathrm{1s}}(s_{\mathrm C})=T_{\mathrm C}\).} &
\parbox[t]{0.21\textwidth}{Used for accumulation maps and size convergence. We use \(T_{\mathrm H}=310\,\mathrm K\), \(T_{\mathrm C}=300\,\mathrm K\), \(w=4a\), and \(s_0=s_{\mathrm H}+3a\). For the \((\beta,\delta)\) scan, \(s_0=s_{\mathrm H}+0.50L\) and \(w=0.30L\), with \(L=s_{\mathrm C}-s_{\mathrm H}\).} \\[1ex]

\parbox[t]{0.12\textwidth}{Center-hot} &
\parbox[t]{0.39\textwidth}{Let \(s_{\mathrm m}=(s_{\mathrm H}+s_{\mathrm C})/2\), \(s_{\mathrm L}=s_{\mathrm m}-W/2\), and \(s_{\mathrm R}=s_{\mathrm m}+W/2\). Define
\(\displaystyle r(s)=\frac12\left[\tanh\!\left(\frac{s-s_{\mathrm L}}{w}\right)-\tanh\!\left(\frac{s-s_{\mathrm R}}{w}\right)\right]\).
Because \(r(s)\) has small nonzero tails at the finite boundaries, define
\(\displaystyle b(s)=r(s_{\mathrm H})+\frac{s-s_{\mathrm H}}{s_{\mathrm C}-s_{\mathrm H}}\left[r(s_{\mathrm C})-r(s_{\mathrm H})\right]\)
and \(\widetilde r(s)=r(s)-b(s)\). Then \(\widetilde r(s_{\mathrm H})=\widetilde r(s_{\mathrm C})=0\). The normalized profile is
\(\displaystyle p_{\mathrm{ch}}(s)=\frac{\widetilde r(s)}{\max_i\widetilde r(s_i)}\)
and
\(\displaystyle T_{\mathrm{ch}}(s)=T_{\mathrm C}+(T_{\mathrm H}-T_{\mathrm C})p_{\mathrm{ch}}(s)\).} &
\parbox[t]{0.21\textwidth}{Used for real-space accumulation and current maps. We use \(T_{\mathrm H}=310\,\mathrm K\), \(T_{\mathrm C}=300\,\mathrm K\), \(W=6a\), and \(w=1.5a\).} \\

\bottomrule
\end{tabular}
\end{table}

\newpage

\section{Symmetry-resolved predictions}

The analytical results give the following consequences for the checkerboard lattice.

The equilibrium checkerboard PAM is linear in both orders at weak coupling,
\begin{equation}
 L_{n\mathbf k,\mathrm{cb}}^z\propto\beta\delta,
\end{equation}
with nodes on $k_x=0$ and $k_y=0$ and opposite signs in adjacent momentum-space quadrants.  Reversing either the N\'eel domain or the $k_2$--$k_3$ pattern reverses the texture.

The thermal splitter coefficient satisfies
\begin{equation}
 D(0,\delta)=D(\beta,0)=0,
 \qquad
 D(-\beta,\delta)=D(\beta,-\delta)=-D(\beta,\delta).
\end{equation}
In the square axes it is a symmetric off-diagonal conductivity.  In the two diagonal axes it becomes the pair of longitudinal conductivities $+D$ and $-D$.  Rotating the temperature gradient by $90^{\circ}$ reverses the transverse splitter current, while a $45^{\circ}$ gradient produces a purely longitudinal checkerboard current.

Uniform diagonal springs, $k_2=k_3$, remove the checkerboard splitter but retain elastic polarization mixing. Removing all diagonal springs also removes this elastic mixing, although gyroscopic coupling can still mix the two polarizations. The PAM Hall response must then be calculated separately.

The checkerboard-projected coherent response reverses sign under either domain reversal, like the thermal splitter. Its frequency dependence follows from Eq.~\eqref{eq:Green}, while Eq.~\eqref{eq:stokes_response} describes its dependence on excitation polarization. The relation between opposite-edge signals also depends on the excitation. Linear, circular, and unpolarized sources probe different combinations of the polarization response.

\appendix

\section{Periodic bond list}
\label{app:bonds}

Table~\ref{tab:bonds} lists one orientation of every undirected bond in the $2a\times2a$ cell.  The translation gives the cell containing the second endpoint relative to the first endpoint.

\begin{table}[h]
\centering
\caption{Periodic bonds of the four-site checkerboard cell.}
\label{tab:bonds}
\begin{tabular}{@{}ccccc@{}}
\toprule
First site & Second site & Translation$/a$ & Direction & Stiffness\\
\midrule
$A$ & $B$ & $(0,0)$ & $(1,0)$ & $k_1$\\
$A$ & $C$ & $(0,0)$ & $(0,1)$ & $k_1$\\
$B$ & $A$ & $(2,0)$ & $(1,0)$ & $k_1$\\
$B$ & $D$ & $(0,0)$ & $(0,1)$ & $k_1$\\
$C$ & $D$ & $(0,0)$ & $(1,0)$ & $k_1$\\
$C$ & $A$ & $(0,2)$ & $(0,1)$ & $k_1$\\
$D$ & $C$ & $(2,0)$ & $(1,0)$ & $k_1$\\
$D$ & $B$ & $(0,2)$ & $(0,1)$ & $k_1$\\
$A$ & $D$ & $(0,0)$ & $(1,1)$ & $k_2$\\
$B$ & $C$ & $(0,0)$ & $(-1,1)$ & $k_2$\\
$C$ & $B$ & $(0,2)$ & $(1,1)$ & $k_3$\\
$D$ & $A$ & $(0,2)$ & $(-1,1)$ & $k_3$\\
$B$ & $C$ & $(2,0)$ & $(1,1)$ & $k_3$\\
$A$ & $D$ & $(-2,0)$ & $(-1,1)$ & $k_3$\\
$D$ & $A$ & $(2,2)$ & $(1,1)$ & $k_2$\\
$C$ & $B$ & $(-2,2)$ & $(-1,1)$ & $k_2$\\
\bottomrule
\end{tabular}
\end{table}

\clearpage

\section{Expanded checkerboard angular-momentum coefficient}
\label{app:Lambda}

For isolated positive-frequency parent-lattice branches, all elastic eigenvectors and their first-order checkerboard corrections can be chosen real. Consequently, the factors of $\ii$ associated with the two real antisymmetric generators $\boldsymbol{\mathcal J}_{\mathrm{AF}}$ and $\boldsymbol{\mathcal J}$ cancel exactly, as in Eq.~\eqref{eq:L_beta}, and no imaginary-part operation is required. Differentiating Eq.~\eqref{eq:Lambda} directly and applying the product rule gives
\begin{align}
 \Lambda_n(\mathbf k)
 &=4\left.\frac{\partial\omega_n}{\partial\delta}\right|_{0}
 \sum_{m\ne n}
 \frac{
 \langle m|\boldsymbol{\mathcal J}_{\mathrm{AF}}|n\rangle
 \langle n|\boldsymbol{\mathcal J}|m\rangle
 }{\omega_n^2-\omega_m^2}
 \nonumber\\
 &\quad+4\omega_n\sum_{m\ne n}
 \frac{1}{\omega_n^2-\omega_m^2}
 \left.\frac{\partial}{\partial\delta}\left[
 \langle m|\boldsymbol{\mathcal J}_{\mathrm{AF}}|n\rangle
 \langle n|\boldsymbol{\mathcal J}|m\rangle
 \right]\right|_{0}
 \nonumber\\
 &\quad-4\omega_n\sum_{m\ne n}
 \frac{
 \langle m|\boldsymbol{\mathcal J}_{\mathrm{AF}}|n\rangle
 \langle n|\boldsymbol{\mathcal J}|m\rangle
 }{(\omega_n^2-\omega_m^2)^2}
 \left[
 2\omega_n\left.\frac{\partial\omega_n}{\partial\delta}\right|_{0}
 -2\omega_m\left.\frac{\partial\omega_m}{\partial\delta}\right|_{0}
 \right].
 \label{eq:Lambda_expanded}
\end{align}
The derivative of the numerator is
\begin{align}
 &\left.\frac{\partial}{\partial\delta}\left[
 \langle m|\boldsymbol{\mathcal J}_{\mathrm{AF}}|n\rangle
 \langle n|\boldsymbol{\mathcal J}|m\rangle
 \right]\right|_{0}
 \nonumber\\
 &\quad=
 \left\langle \left.\frac{\partial m}{\partial\delta}\right|_{0}\right|
 \boldsymbol{\mathcal J}_{\mathrm{AF}}|n\rangle
 \langle n|\boldsymbol{\mathcal J}|m\rangle
 +\langle m|\boldsymbol{\mathcal J}_{\mathrm{AF}}
 \left|\left.\frac{\partial n}{\partial\delta}\right|_{0}\right\rangle
 \langle n|\boldsymbol{\mathcal J}|m\rangle
 \nonumber\\
 &\qquad+
 \langle m|\boldsymbol{\mathcal J}_{\mathrm{AF}}|n\rangle
 \left\langle \left.\frac{\partial n}{\partial\delta}\right|_{0}\right|
 \boldsymbol{\mathcal J}|m\rangle
 +\langle m|\boldsymbol{\mathcal J}_{\mathrm{AF}}|n\rangle
 \langle n|\boldsymbol{\mathcal J}
 \left|\left.\frac{\partial m}{\partial\delta}\right|_{0}\right\rangle .
 \label{eq:Lambda_numerator_derivative}
\end{align}
The frequency and eigenvector derivatives are those given in Eqs.~\eqref{eq:omega_delta} and \eqref{eq:state_delta}. Substituting those expressions into Eqs.~\eqref{eq:Lambda_expanded} and \eqref{eq:Lambda_numerator_derivative} yields the corresponding triple-band expression involving only parent-lattice eigenvalues, eigenvectors, and the matrix derivative in Eqs.~\eqref{eq:Ddelta_AA}--\eqref{eq:Ddelta_off}. At a band degeneracy, the checkerboard perturbation is first diagonalized within the complete degenerate subspace before these nondegenerate formulas are applied to states outside that subspace.

\begingroup
\makeatletter
\let\@FMN@list\@empty
\let\saved@label\label
\def\label#1{%
  \def\temp@label{#1}\def\last@bib{LastBibItem}%
  \ifx\temp@label\last@bib\else\saved@label{#1}\fi
}
\makeatother

\endgroup

\end{document}